\documentclass[acmsmall]{acmart}
\usepackage[utf8]{inputenc}
\usepackage[T1]{fontenc}
\usepackage{microtype}
\usepackage[ruled,linesnumbered, noend]{algorithm2e}
\usepackage{listings}
\usepackage{makecell}
\usepackage{longfbox}
\usepackage{subcaption}
\usepackage{xcolor}
\usepackage{enumitem}
\usepackage{algpseudocode}
\usepackage{multirow}
\usepackage{cleveref} 
\usepackage{siunitx}

\creflabelformat{equation}{#2 #1#3}

\lstdefinestyle{cpp}{
  language=C++,
  basicstyle=\ttfamily\normalsize,
  keywordstyle=\color{blue},
  stringstyle=\color{red},
  commentstyle=\color{green!50!black},
  morecomment=[l][\color{green!50!black}]{//},  
  morecomment=[s][\bfseries\color{red}]{/*}{*/},  
  numbers=left,
  numberstyle=\tiny\color{gray},
  stepnumber=1,
  numbersep=5pt,
  backgroundcolor=\color{gray!10},
  frame=single,
  rulecolor=\color{black},
  tabsize=2,
  captionpos=b,
  breaklines=false,
  breakatwhitespace=false,
  showspaces=false,
  showstringspaces=false,
  showtabs=false,
  morekeywords={},
  linewidth=0.98\linewidth
}

\definecolor{ap-traj-yellow}{RGB}{213, 211, 118}
\definecolor{ap-traj-green}{RGB}{162, 245, 122}
\definecolor{ap-traj-purple}{RGB}{235, 51, 176}
\definecolor{ap-traj-cyan}{RGB}{93, 158, 170}
\definecolor{ap-traj-gray}{RGB}{222, 230, 234}
\definecolor{ap-traj-red}{RGB}{251, 29, 29}
\definecolor{ap-gt-pink}{RGB}{255,207,234}
\definecolor{ap-gt-light-blue}{RGB}{224, 255, 255}
\definecolor{ap-gt-blue}{RGB}{91,155,213}
\definecolor{ap-gt-red}{RGB}{255,0,0}

\let\oldnl\nl
\newcommand{\nlnonumber}{\renewcommand{\nl}{\let\nl\oldnl}}

\definecolor{revisioncolor}{RGB}{0,0,0} 

\def\prjName{RouthSearch}

\setcopyright{cc}
\setcctype{by}
\acmDOI{10.1145/3728904}
\acmYear{2025}
\acmJournal{PACMSE}
\acmVolume{2}
\acmNumber{ISSTA}
\acmArticle{ISSTA029}
\acmMonth{7}
\received{2024-10-31}
\received[accepted]{2025-03-31}

\begin{document}

\title{RouthSearch: Inferring PID Parameter Specification for Flight Control Program by Coordinate Search}

\author{Siao Wang}
\orcid{0009-0008-1807-2746}
\affiliation{%
  \institution{Fudan University}
  \city{Shanghai}
  \country{China}
}
\email{22110240039@m.fudan.edu.cn}

\author{Zhen Dong}
\authornote{Corresponding author}
\orcid{0009-0009-1193-0696}
\affiliation{%
  \institution{Fudan University}
  \city{Shanghai}
  \country{China}
}
\email{zhendong@fudan.edu.cn}

\author{Hui Li}
\orcid{0009-0009-5826-343X}
\affiliation{%
  \institution{Fudan University}
  \city{Shanghai}
  \country{China}
}
\email{22210240023@m.fudan.edu.cn}

\author{Liwei Shen}
\orcid{0000-0002-8107-0590}
\affiliation{%
  \institution{Fudan University}
  \city{Shanghai}
  \country{China}
}
\email{shenliwei@fudan.edu.cn}

\author{Xin Peng}
\orcid{0000-0003-3376-2581}
\affiliation{%
  \institution{Fudan University}
  \city{Shanghai}
  \country{China}
}
\email{pengxin@fudan.edu.cn}

\author{Dongdong She}
\orcid{0000-0001-6655-0468}
\affiliation{%
  \institution{Hong Kong University of Science and Technology}
  \city{Hong Kong}
  \country{China}
}
\email{dongdong@cse.ust.hk}

\begin{abstract}
Flight control programs are widely used in unmanned aerial vehicles (UAVs) to manage and maintain UAVs' flying behaviors dynamically. These flight control programs include a PID control module that takes three user-configurable PID parameters: Proportional (P), Integral (I), and Derivative (D). Users can also adjust these PID parameters during flight to suit the needs of various flight tasks. However, flight control programs do not have sufficient safety checks on the user-provided PID parameters, leading to a severe vulnerability of UAV---input validation bug. It happens when the user misconfigures PID parameters and causes the UAV to enter a dangerous state, such as deviation from the expected path, loss of control, or even crash. 

Prior works use random testing approaches like fuzzing to identify invalid PID parameters from user input. However, they are not effective in the three-dimensional search space of PID parameters. Meanwhile, each dynamic execution of the UAV test is very expensive, further affecting the performance of random testing.

In this work, we address the problem of PID parameter misconfiguration by combining the Routh-Hurwitz stability criterion with coordinate search, introducing a method called \prjName. Instead of identifying misconfigured PID parameters in an ad-hoc fashion, \prjName\ principledly determines valid ranges for three-dimensional PID parameters. We first leverage the Routh-Hurwitz Criterion to identify a theoretical PID parameter boundary. We then refine the boundary using an efficient coordinate search. The valid range of three-dimensional PID parameters determined by \prjName\ can filter out misconfigured PID parameters from users during flight and further help to discover logical bugs in popular flight control programs. 

We evaluated \prjName\ across eight flight modes in two popular flight control programs, PX4 and ArduPilot. The results show that \prjName\ can determine the valid ranges of the three-dimensional PID parameters with an accuracy of 92. 0\% when compared to the ground truth. In terms of the total number of misconfigured PID parameters, \prjName\ discovers 3,853 sets of PID misconfigurations within 48 hours, while the STOA work PGFuzz only discovers 449 sets of PID misconfigurations, significantly outperforming prior works by 8.58 times. Additionally, our method helps to detect three bugs in ArduPilot and PX4.
\end{abstract}




\maketitle

\section{Introduction}
\label{sec:introduction}
Unmanned Aerial Vehicles (UAVs) are widely used in various domains, such as agriculture \cite{uav-agriculture}, meteorology \cite{uav-meteorology}, search and rescue operations \cite{uav-search-rescue}, and modern warfare \cite{military-uav}.
They rely on flight control programs to manage the UAV's position and attitude during flight. 
Flight control programs include a PID control module with three ~\emph{user-configurable} PID parameters: Proportional (P), Integral (I), and Derivative (D). 
The PID parameters can be dynamically adjusted in flight to meet the requirements of various UAV tasks. However, there are no sufficient safety checks on the PID parameters before they are dynamically adjusted~\citep{han2022control,han2024range}. The lack of configuration safety checks on the PID parameters leads to a serious security problem in the UAV flight control program, named \emph{input validation bug}~\cite{kim2019rvfuzzer}.
The input validation bug occurs when a user passes a set of misconfigured PID parameters to the flight control program, further causing the UAV to turn into a dangerous flight state, such as deviations from the intended flight paths, collisions, lost control, and crashes~\citep{24KGQuadLight,  ArduPilot2024crash, massie2022amazon, atsb_docklands_drone_incident}.
As flight control programs like PX4 \cite{px4-website} and ArduPilot \cite{ArduPilot-website} are widely used in UAVs, the input validation bug is becoming a key problem in UAV safety. A protection technique that can identify dynamically misconfigured PID parameters during flight is urgently needed. 

Prior work PGFuzz \cite{kim2021pgfuzz} uses random fuzzing to detect  PID parameters misconfiguration in an ad-hoc fashion. RVFuzzer \cite{kim2019rvfuzzer} and LGDFuzzer \cite{han2022control} identify valid ranges for each individual PID parameter through binary search and machine learning-based prediction, respectively. 
However, these approaches treat each parameter independently, overlooking their interdependencies and thus \textcolor{revisioncolor}{are ineffective in identifying} valid ranges when taking these relationships into account.

Identifying valid ranges for PID parameters is challenging. First, the classic stability model that can reason the validity of PID parameters is not directly applicable in real-world scenarios. The stability theory only reasons the mathematical model of the PID control and gives a theoretical boundary that separates the valid and invalid PID parameters~\cite {castillo2021preliminary, aastrom1989towards}. In practice, the flying behavior of UAVs can be influenced by many factors other than the underlying mathematical model, such as the implementation of the flight control program and environmental noise. These factors cause a \emph{drastic shift} on its theoretical PID parameter boundary from the real one. 
Second, the search space for PID parameters is three-dimensional and includes millions of possible parameter sets. The runtime cost to validate each set of PID parameters is very high, taking hundreds of seconds to run on a UAV simulator. 

To address these challenges, an efficient dynamic technique is desired because it is lightweight and resistant to unknown physical noise.
We observe that despite the gap between the theoretical PID parameter boundary and the real one being unknown, the real PID parameter boundary is continuous and fluctuates around the theoretical one. Therefore, we start from the theoretical boundary, then use the coordinate search~\cite{wright2015coordinate} to quickly depict such a gap and construct an accurate PID parameter boundary.  

In this paper, we propose a principled approach called \prjName\ to efficiently determine the valid range for the three-dimensional PID parameters. Given these valid ranges, we can detect PID misconfigurations of UAVs on the fly. Specifically, we first leverage the Routh-Hurwitz  criterion~\cite{ho1998elementary} from stability theory to derive a theoretical PID parameter boundary as the starting line. Then, we perform an efficient coordinate search to quickly refine the boundary by identifying the unknown boundary shift. To detect whether a PID parameter configuration is valid, we implement a UAV \textcolor{revisioncolor}{misbehavior validator} as an oracle. We record the UAV flight log and analyze it against Metric Temporal Logic encoding the UAV specification. If a UAV's flight behavior violates the MTL specification, we classify the configuration as invalid.
\prjName\ can identify valid ranges of PID control parameters in different modes defined in mainstream flight control program. \prjName\ also helps users avoid misconfiguring PID parameters and even discover logic bugs in implementing a widely adopted flight control program, ArduPilot.

We evaluated \prjName\ under the eight typical flight modes of two popular flight control programs (i.e., PX4 and ArduPilot).
We measure the performance of \prjName\ by computing the ratio of true PID parameter misconfigurations out of all predicted PID parameter misconfigurations. Our experimental results show that \prjName\ can identify the valid range of PID parameters with an accuracy of 92.0\%. We also evaluate \prjName's ability to discover individual PID parameter misconfigurations. The result shows that \prjName\ discovers an average of 3,853 misconfigurations within 48 hours, while the baseline tool PGFuzz has only discovered 449 misconfigurations.

Our key contributions are summarized as follows:
\begin{itemize}
  \item We propose a principled method to identify the valid range of three-dimensional PID parameters. The valid range can help prevent users from misconfiguring PID parameters during flight.

  \item We use an efficient coordinate search to identify individual PID parameter misconfigurations on the UAV flight control program.

  \item We evaluate our method under 8 flight modes on two popular open-source flight control programs. The evaluation result shows that \prjName\ identifies the boundary of valid PID parameter regions with 92.0\% accuracy. 

  \item We open-source our tool to help users properly set up the PID parameters and further foster future research in this domain.
  We have released our implementation in the web at \href{https://github.com/SciC0d3m4xOfW/RouthSearch}{https://github.com/SciC0d3m4xOfW/RouthSearch}.
\end{itemize}

The rest of the paper is organized as follows. \Cref{sec:background} summarizes the background of the UAV PID control and the Routh-Hurwitz
Stability Criterion. \Cref{sec:problem_analysis} explains the challenges of identifying valid ranges for PID parameters and the insights behind our method. \Cref{sec:methodology} introduces our methodology in detail.
We present our experimental results in \Cref{sec:evaluation} and discuss the limitations of our method in \Cref{sec:limitation_and_discussion}. We conclude with a summary of related work in \Cref{sec:related_works} and a conclusion in \Cref{sec:conclusion}.

\section{Background}
\label{sec:background}
\subsection{PID Control Module}
 
The PID control module regulates the UAV's position and attitude to maintain its control and stability.
It consists of two sub-modules: 1) position sub-module to manage UAV's position and velocity; 2) attitude sub-module to manage UAV's attitude and the angular velocity \citep{lopez2023pid, mottola2018fundamental}. 
The PID control module has three user-configurable PID parameters, corresponding to Proportional (P), Integral (I), and Derivative (D). The three user-configured PID parameters are used to dynamically maintain the UAV's stability during flight.

\begin{figure}[htbp]
  \centering
  \begin{subfigure}[m]{0.2\linewidth}
    \centering
    \includegraphics[width=\linewidth]{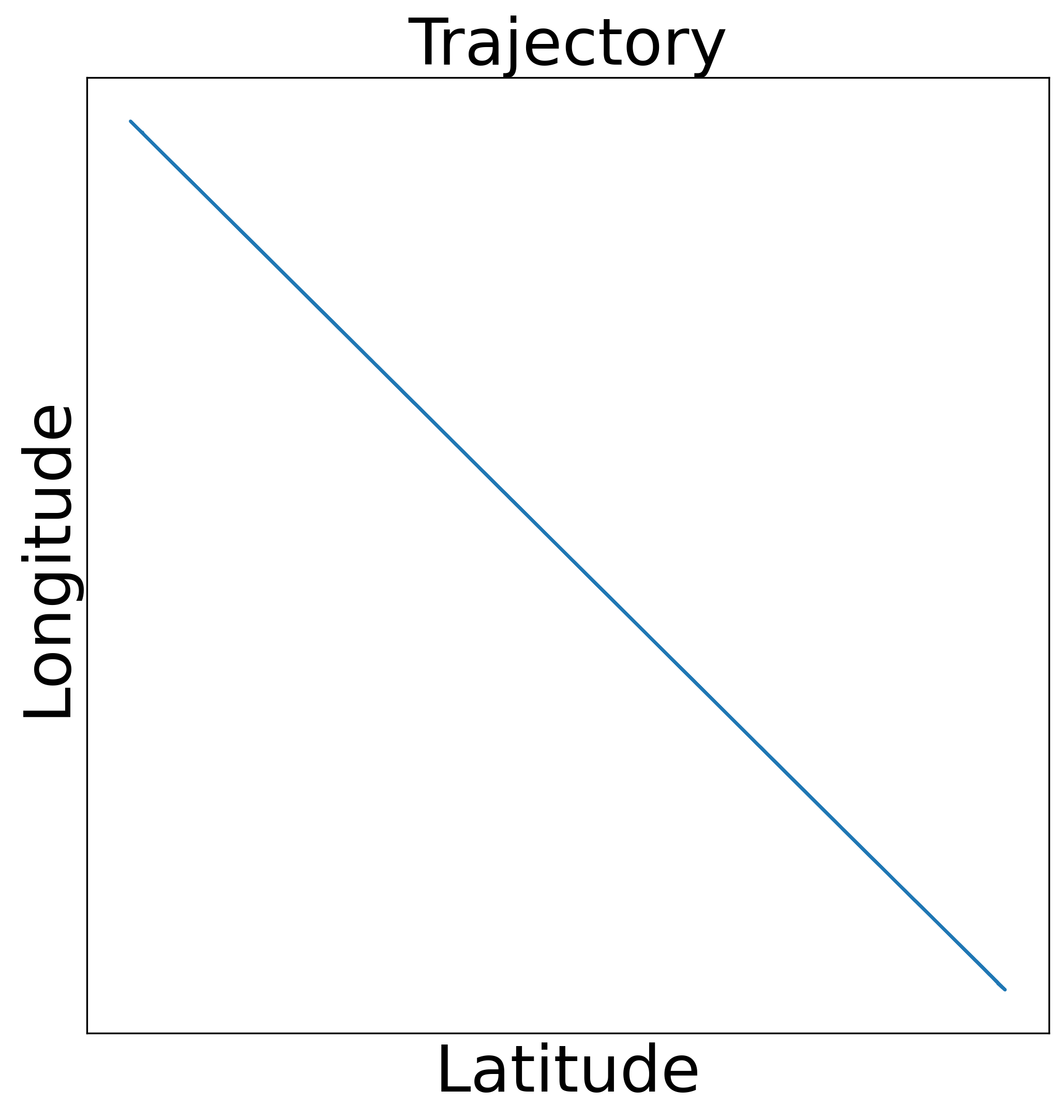} 
    \captionsetup{
    labelfont={color=revisioncolor, sf, small},
    textfont={color=revisioncolor, sf, small} 
    }
    \caption{Valid PID Config}
    \label{fig:traj_before_invalid_pid}
  \end{subfigure}
  \hspace{.1\linewidth} 
  \begin{subfigure}[m]{0.2\linewidth}
    \centering
    \includegraphics[width=\linewidth]{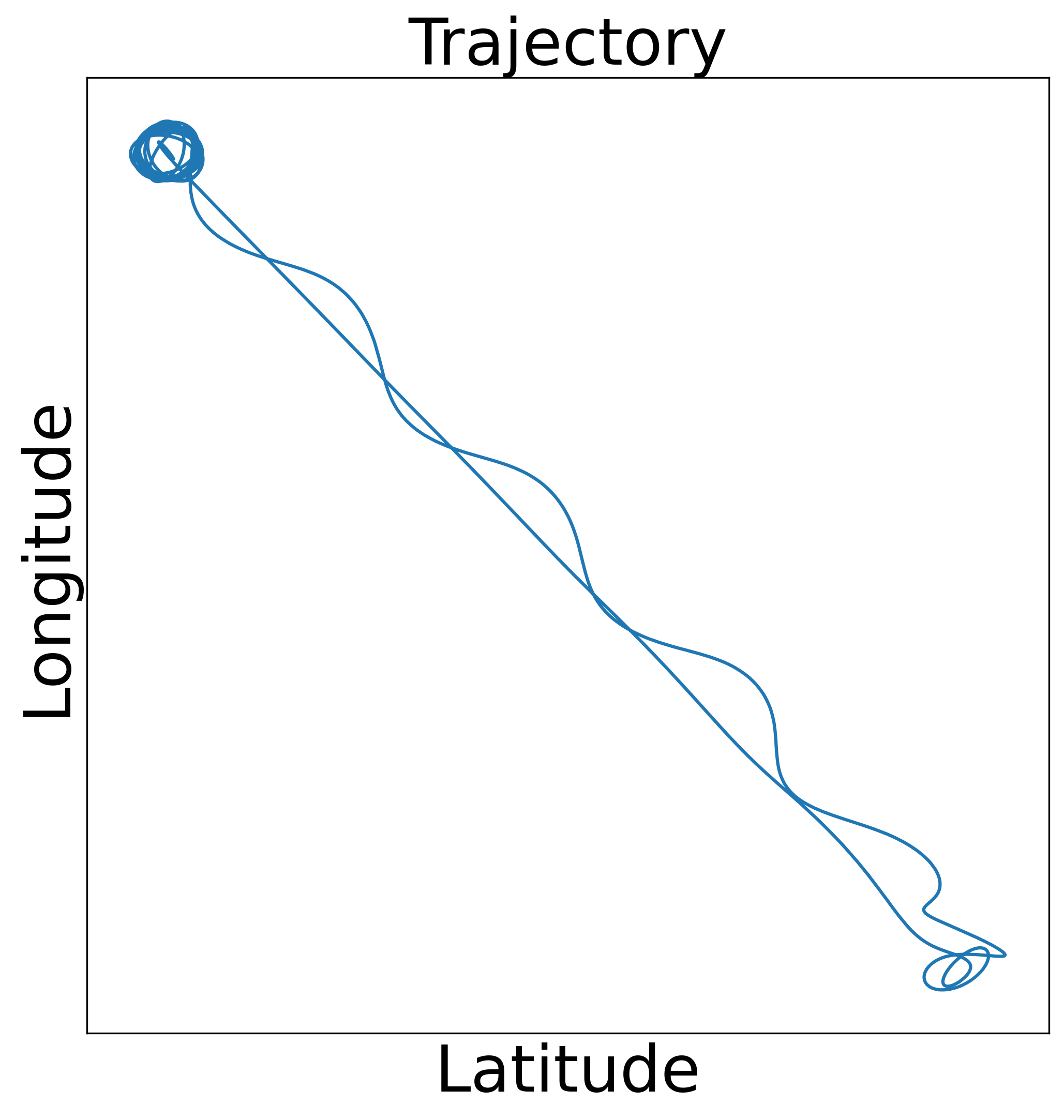} 
    \captionsetup{
    labelfont={color=revisioncolor, sf}, 
    textfont={color=revisioncolor, sf} 
    }
    \caption{{{Invalid PID Config}}}
    \label{fig:traj_after_invalid_pid}
  \end{subfigure}

  \captionsetup{
    labelfont={color=revisioncolor, sf, small}, 
    textfont={color=revisioncolor, sf, small} 
  }  
  \Description{Comparison of UAV trajectories before and after setting invalid PID configurations. After the invalid PID parameter is set, the UAV trajectory becomes waving and unstable as shown in \Cref{fig:traj_after_invalid_pid}.}
  \caption{Comparison of UAV trajectories before and after setting invalid PID configurations. After the invalid PID parameter is set, the UAV trajectory becomes waving and unstable as shown in \Cref{fig:traj_after_invalid_pid}.}
  \label{fig:trajectory_comparison_invalid_pid}
\end{figure}

The error in a control system is typically defined as the difference between the desired state and the actual state of the system. A control system is stable when its error approaches zero as time approaches infinity. A stable UAV indicates that both the position and attitude sub-modules are properly configured and have around zero errors \cite{fantoni2008asymptotic}. While misconfigured PID parameters can lead to large errors and an unstable UAV (e.g., oscillating or diverging from the expected path).

\textcolor{revisioncolor}{To illustrate the significant impact of PID parameter configurations, \Cref{fig:trajectory_comparison_invalid_pid} provides an example of UAV trajectory under valid/invalid PID configurations. \Cref{fig:traj_before_invalid_pid} demonstrates a stable flight trajectory from the bottom-right corner to the top-left corner and back. While \Cref{fig:traj_after_invalid_pid} depicts an unstable flight trajectory, characterized by unstable and waving movements and localized circular patterns, which highlight the instability of the UAV resulting from invalid PID configurations.}

\subsection{Routh-Hurwitz Stability Criterion} 
The Routh-Hurwitz Criterion is a mathematical condition used to determine the stability of a control system \cite{ho1998elementary}.
For a closed-loop PID control system, we can have the following equation:
{\small
\begin{equation}
    \label{eq:uav-dynamic-model}
    \frac{\mathrm{d}^2 x(t)}{\mathrm{d} t^2} + a_2\frac{\mathrm{d} x(t)}{\mathrm{d} t} + a_1x(t) = u(t) 
\end{equation}
\begin{equation}
    u(t) = k_p e(t) + k_i \int_{0}^{t} e(\tau) \, d\tau + k_d \frac{\mathrm{d} e(t)}{\mathrm{d} t}
\end{equation}
}
where $x(t)$ and $u(t)$ denote the system output and the control input at time step $t$ respectively. $e(t)$ represents the error at time $t$. 
$a_1$ and $a_2$ are constant coefficients set by the system. $k_p$, $k_i$, and $k_d$ are the PID control's proportional, integral, and derivative parameters. According to the Routh-Hurwitz Criterion, when the following condition is satisfied: 
{\small
\begin{equation}
\label{Routh-Hurwitz:criterion1}
k_p + a_1 > 0  
\end{equation}
\begin{equation}
\label{Routh-Hurwitz:criterion2}
k_d + a_2 > 0  
\end{equation}
\begin{equation}
\label{Routh-Hurwitz:criterion3}
k_i > 0  
\end{equation}
\begin{equation}
\label{Routh-Hurwitz:criterion4}
(k_p+a_1)(k_d+a_2)> k_i 
\end{equation}
}
the system is stable \citep{ho1998elementary, zhao2017pid}. $k_p$, $k_i$, and $k_d$ are manually configured by users. Due to a lack of sufficient safety checks on these PID parameters in flight, incorrect configurations by users can lead to severe consequences such as deviation from the expected path, collisions, loss of control, and crashes. Note that the Routh-Hurwitz Criterion is a theoretical model of a low-level control system and hence cannot be directly applied to UAV systems. There exists a gap between the theoretical boundary derived by the Routh-Hurwitz Criterion and the real one. We will explain the root cause of such boundary shifts in the following section. 

\section{Challenges and Insight}
\label{sec:problem_analysis}
Routh-Hurwitz stability analysis shows that there exists a boundary that separates valid and invalid PID parameter configurations~\cite{zhao2017pid}. However, we cannot directly use the Routh-Hurwitz Criterion to derive a valid PID parameter boundary due to an unknown gap between the theoretical and real boundaries. We first explain three challenges in determining a valid PID parameter boundary. We then introduce our insights using an efficient coordinate search to quickly depict the unknown gap and construct the valid PID parameter boundary. 

\noindent \textbf{Challenge: Flight Mode Influence.} Flight mode can greatly influence the validity of the PID parameter. Popular flight control program defines various flight modes. Each mode represents a unique flight behavior that imposes corresponding constraints on the PID control module. A stable PID parameter configuration satisfying the Routh-Hurwitz Criterion may fail to meet these additional constraints defined in flight mode, resulting in unexpected flying behaviors or failed flight missions. Hence, there is no silver bullet PID parameter set that is valid across diverse flight modes. A valid PID parameter range must be \emph{mode-specific}. 
For example, a PID parameter configuration $(p, i, d)$ satisfies the Routh-Hurwitz Criterion, and hence, the UAV configured with $(p, i, d)$ can fly stably. However, under a certain mode, the UAV configured with $(p, i, d)$ could fail to accomplish the required flying behaviors and lead to a crash \cite{ArduPilot_autotune_discussion}. 

\noindent \textbf{Challenge: External Noise.} External noises in sensors and environments contribute random variance to the valid PID parameter range. PID control modules are easily affected by high-frequency noise during a flying mission. Even a slight disturbance from sensor input or the environment can cause a significant oscillation in the UAV \cite{schuhmann2011improving}.
Although the developers have spent a considerable amount of time modeling and performing stability analysis of a UAV to derive a valid PID parameter range, the UAV may still lose control or crash due to improper PID parameter configuration.
Consequently, the PID parameter range guidance in existing documentation is prone to erroneous. As we can see from \Cref{subsec:Motivated_Example}, a PID parameter configuration falling in the suggested parameter ranges in the document can still cause an unexpected UAV malfunction.

\noindent \textbf{Challenge: Huge PID Parameter Space.} Given the unknown gap between the theoretical and real boundaries, an ideal way to determine the real boundary is to use a dynamic testing approach that can adapt to diverse flight modes and random noise. However, the PID parameter space is a three-dimensional space, composed of a huge number of possible PID parameter configurations. For instance, the search space of a popular UAV flight control program, ArduPilot, is $\textmd{[0.1, 6]}$, $\textmd{[0.02, 1]}$, and $\textmd{[0.0, 1]}$. 
We set the search step size for each dimension as $\textmd{0.1, 0.01,}$ and $\textmd{0.001}$, respectively. There are a total of around 6 million parameter sets. Meanwhile, the runtime cost to verify the validity of each PID parameter configuration is very high, either through a real UAV device or a simulator. 

\noindent \textbf{Insight: Coordinate Search.} 
Coordinate search is an efficient search algorithm that can perform a coordinate-wise search to minimize a specified objective function or other search goals. Empirically, we observe that although the unknown gap exists between the theoretical and real boundaries of valid PID parameter configurations, the real boundary remains continuous and oscillates around the theoretical one. Therefore, we can leverage the coordinate search algorithm to depict the unknown gap with an efficient dynamic search. We start from the theoretical boundary derived by the Routh-Hurwitz Criterion and refine it with the gap identified by the coordinate search. In the end, we can efficiently construct the valid PID parameter boundary. 

\section{Methodology}
\label{sec:methodology}
\subsection{Overview}
\begin{figure}[!htbp]
    \centering
    \includegraphics[width=0.6\linewidth]{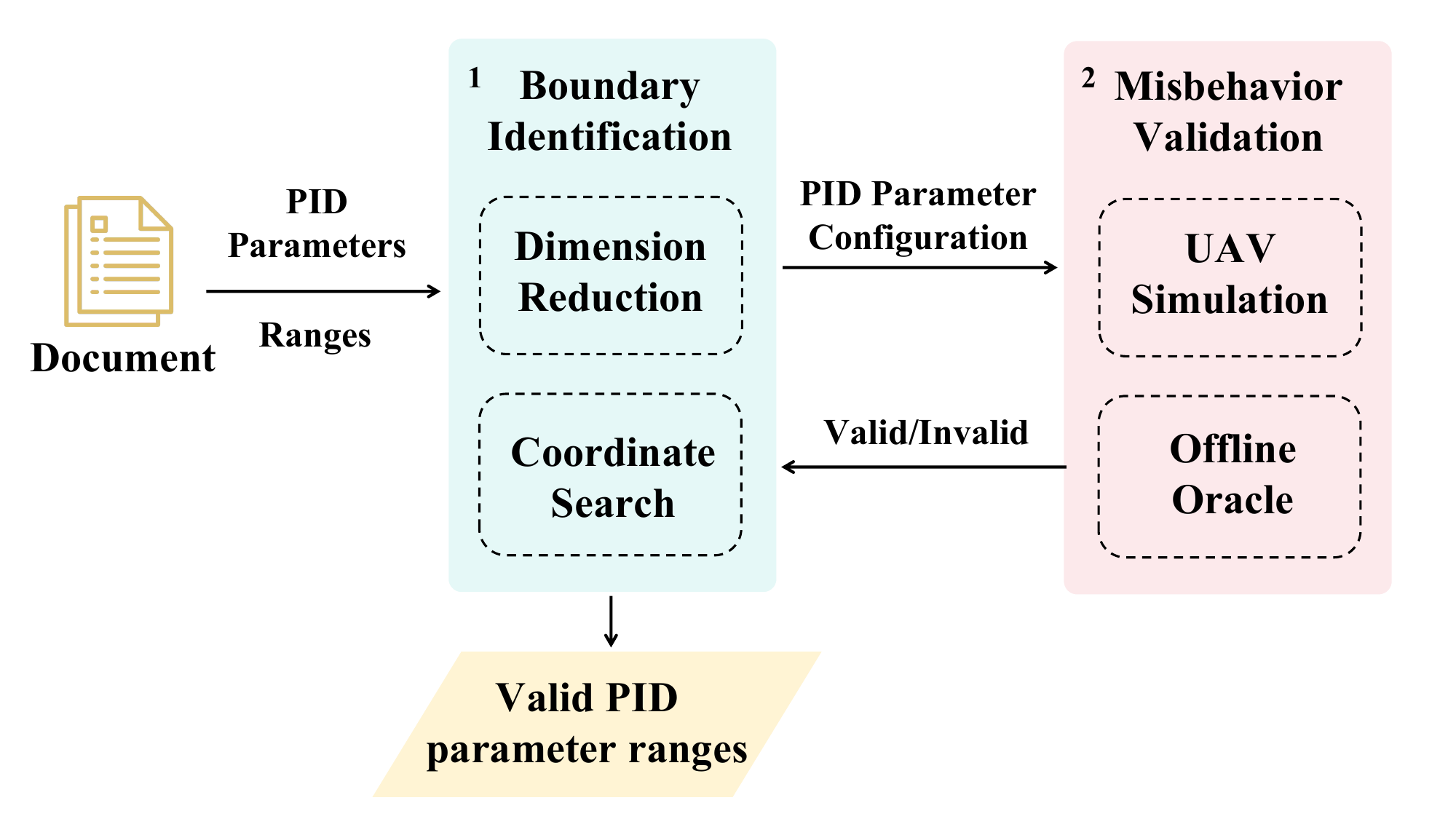}
    \Description{\prjName\ Overview: \prjName\ includes Boundary Identification module and \textcolor{revisioncolor}{Misbehavior Validation} module. The Boundary Identification module takes in PID parameters along with their ranges from the document and then explores the parameter space to find boundaries between valid and invalid configurations. The Boundary Identification module queries the \textcolor{revisioncolor}{Misbehavior Validation} module to validate PID parameter configurations during the search process. Finally, it outputs valid PID parameter ranges.}
    \caption{\prjName\ Overview: \prjName\ includes Boundary Identification module and \textcolor{revisioncolor}{Misbehavior Validation} module. The Boundary Identification module takes in PID parameters along with their ranges from the document and then explores the parameter space to find boundaries between valid and invalid configurations. The Boundary Identification module queries the \textcolor{revisioncolor}{Misbehavior Validation} module to validate PID parameter configurations during the search process. Finally, it outputs valid PID parameter ranges.}
    \label{fig:overview}
\end{figure}
The \prjName\ consists of two components: (1) Boundary Identification and (2) \textcolor{revisioncolor}{Misbehavior Validation}, as illustrated in \Cref{fig:overview}. 
The boundary identification module first takes in PID parameters and their corresponding value range from the official documents of UAV flight control programs. It then efficiently searches the PID parameter space to find a classification boundary between valid and invalid configurations. During the search, the boundary identification module continuously queries the \textcolor{revisioncolor}{misbehavior validation} module to determine whether a given PID parameter configuration is valid. In the end, it outputs the classification boundary.

\subsection{Boundary Identification} \label{sec:boundary_identification}
In this section, we explain the efficient search algorithm that can identify the classification boundary within a three-dimensional PID parameter space. Our approach can generalize to diverse modes of mainstream flight control programs with external noise. 

\noindent \textbf{Dimension Reduction.} 
We analyze the flight control program documentation to extract PID parameters, their corresponding value ranges, and incremental step sizes. We define the three-dimensional PID parameter space as $(k_p, k_i, k_d) \in [p_{\min}, p_{\max}] \times [i_{\min}, i_{\max}] \times [d_{\min}, d_{\max}]$, where $k_p$, $k_i$, and $k_d$ denotes three PID parameter values. $p_{\min}$, $i_{\min}$, and $d_{\min}$ denotes lower bound of PID parameters. Similarly, $p_{\max}$, $i_{\max}$, and $d_{\max}$ denotes the upper bound of PID parameters. We represent the incremental step size as $step_{p}$, $step_{i}$, and $step_{d}$. 
We first divide the three-dimensional PID parameter space $(k_p, k_i, k_d)$ into $n$ two-dimensional planes $(k_i, k_d)$. Each plane has a unique $k_p$ value. The goal of such dimension reduction is that we can run our efficient coordinate search procedure as shown in Line 14-23 in \Cref{algo:boundary_draw} at each two-dimensional plane in $O(n)$. Hence, the total runtime overhead of our approach is $O(n^2)$. While a brutal-force approach has to enumerate all the possible cases within the three-dimensional PID parameter space to find an accurate boundary, whose runtime overhead is $O(n^3)$. 

\begin{figure}[!ht]
  \centering
    \centering
  \begin{subfigure}[m]{0.25\linewidth}
    \centering
    \includegraphics[width=\linewidth]{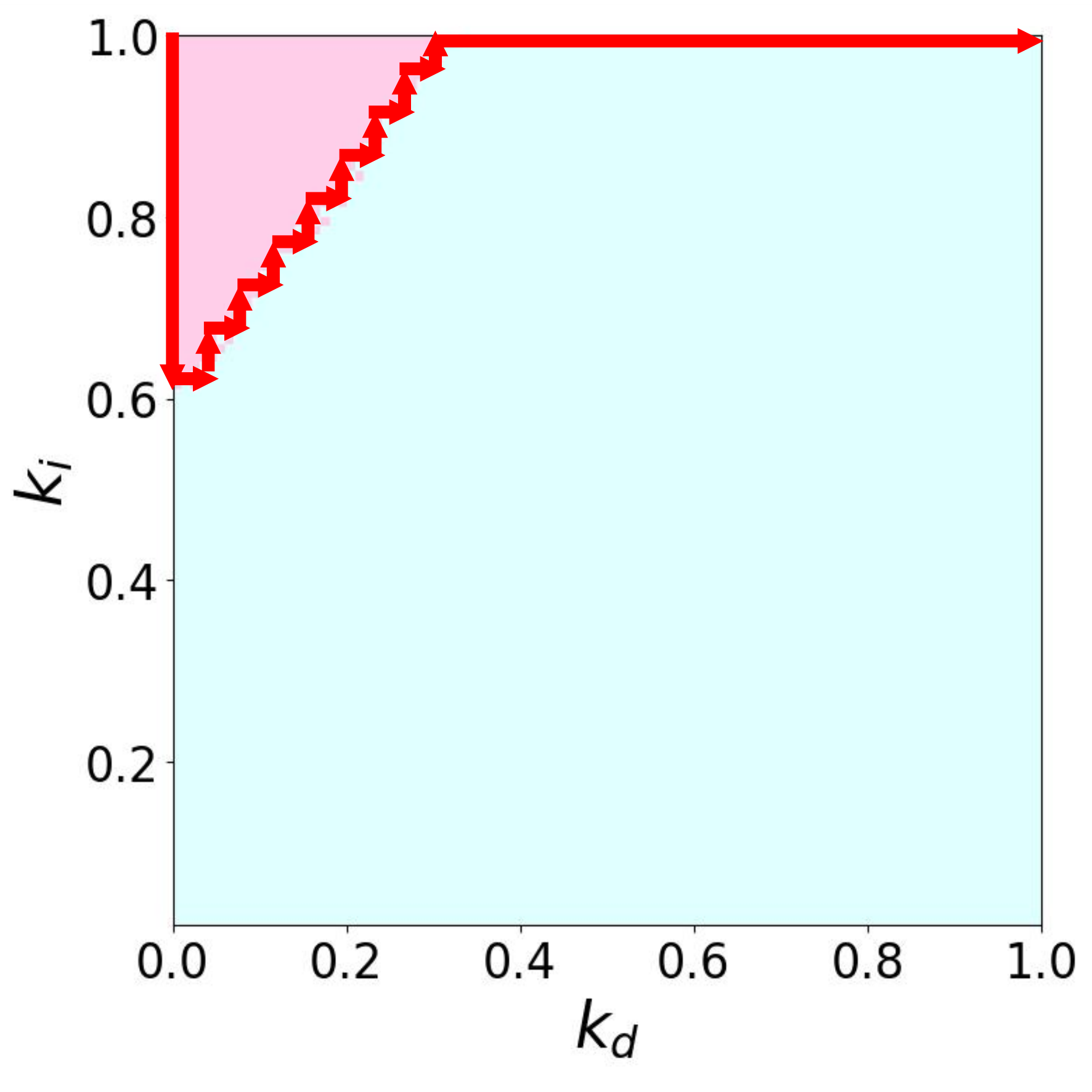} 
    \caption{Linear Boundary Line}
    \label{fig:ArduPilot_boundary_identification_illustration}
  \end{subfigure}
  \hspace{0.1\linewidth}
  \begin{subfigure}[m]{0.275\linewidth}
    \centering
    \includegraphics[width=0.95\linewidth]{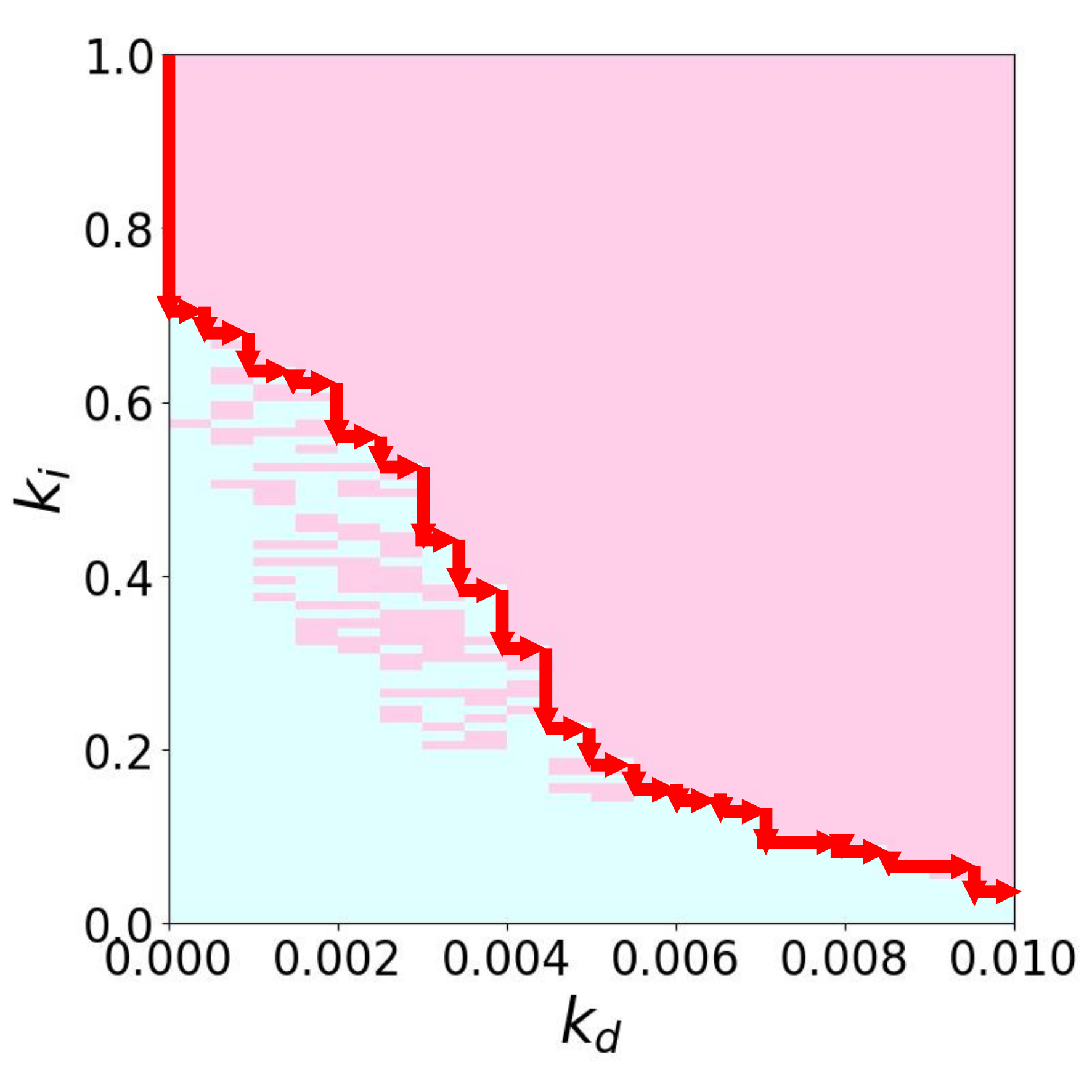}
    \caption{{Noisy Boundary Line}}
    \label{fig:Noise_Illustration}
  \end{subfigure}
  \Description{Different PID parameter boundaries of UAVs. The \colorbox{ap-gt-light-blue}{light blue} and \colorbox{ap-gt-pink}{pink} represent valid and invalid PID configurations, respectively. We highlight the boundary line in \textbf{\textcolor{ap-gt-red}{red}}.}
  \caption{Different PID parameter boundaries of UAVs. The \colorbox{ap-gt-light-blue}{light blue} and \colorbox{ap-gt-pink}{pink} represent valid and invalid PID configurations, respectively. We highlight the boundary line in \textbf{\textcolor{ap-gt-red}{red}}.}
  \label{fig:boundary_identification_illustration}
\end{figure}

As shown in \Cref{Routh-Hurwitz:criterion4}, the Routh-Hurwitz Criterion shows that a classification boundary separating the stable PID parameter values and unstable PID parameter values exists. On a plane $(k_i, k_d)$ with $k_p = p_{const}$, we can have a linear boundary as follows:
{\small
\begin{equation}
\label{eq:linear_cond}
    (p_{const}+a_1)(k_d+a_2) = k_i 
\end{equation}
}
Note that $a_1$ and $a_2$ are unknown constants determined by the physical system and the specific flight mode. 
Ideally, we only need two data points of $(k_i, k_d)$ to determine the value of $a_1$ and $a_2$, 
and further identify the linear boundary. However, real-world UAVs often have \emph{complex non-linear} boundaries due to the unknown boundary shift (see \Cref{sec:problem_analysis}). ~\Cref{fig:Noise_Illustration} shows a noisy boundary of a control system highlighted in \textcolor{red}{red}. 
We can see that the boundary is a messy sawtooth wave rather than a clear linear boundary, while the valid/invalid parameter region remains approximately continuous with no disjoint regions observed empirically. 
To mitigate this problem, we design an efficient algorithm based on coordinate search that is robust to the boundary shift.  

\begin{algorithm}[!ht]
\footnotesize
\nlnonumber
\Indm
\hspace{1cm}
\setlength{\tabcolsep}{3pt}
\begin{tabular}{|l p{4.75in}|}\hline
\textbf{Input}:
    & PID range: $(p_{min}, p_{max}), (i_{min}, i_{max}), (d_{min}, d_{max})$ \\
    & Incremental step: $\{step_p, step_i, step_d\}$ \\
    & Flight Mission: $M$ \\
\textbf{Output}: 
    & Classification boundary: \\
    & $BL$ $=$ 
    $\{(p_1, i_1, d_1), (p_2, i_2, d_2), \cdots, (p_n, i_n, d_n)\}$ \\

\hline
\end{tabular}
\\
\Indp
\SetAlgoNoLine    
\LinesNumbered
\SetKw{KwBy}{by}
\textcolor{purple}{{/* Search dimension $I$ to find a valid config that is nearest to the partition line. */}} \\
\SetKwProg{Fn}{Func Search}{:}{}
\Fn{($p$, $i$, $d$, $direction$, $step_i$)}{
    $is\_valid = False$ \\
    \While{$is\_valid == False$ \textbf{and} $i$ is valid} {
        $is\_valid = $ \textsf{Oracle}$(p, i, d, M)$ \textcolor{purple}\\
        
        \If {$direction == down$} {
            $update = -step_i$ \textcolor{purple}{\Comment{Search downwards}}\\
        } \Else {
            $update = step_i$ \textcolor{purple}{\Comment{Search upwards}}\\
        }
        $i = i + update$  
    }
    $i_{save} = i - update$ \textcolor{purple}{\Comment{Restore $i$ to last valid config}}\\
    \Return $(p, i_{save}, d)$
}

\textcolor{purple}{{ /* Dimension reduction: first fix $P$, then search $(I, D)$ */}}\\
\For{$p$ = $p_{min}$ \KwTo $p_{max}$ \KwBy $step_p$} {
    \textcolor{purple}{ /* Coordinate search over plane $(I, D)$ */} \\
    $i = i_{max}$   \textcolor{purple}{\Comment{Start from upper-left corner}}\\
    \For {$d = d_{min}$ \KwTo $d_{max}$ \KwBy $step_d$} {
        \If {$\textsf{Oracle}(p, i, d, M) == False$ } {
            $(p, i_{save}, d) = $ \textsf{Search}$(p, i, d, down, step_i)$ \textcolor{purple}{\Comment{ Search dimension $i$ downwards}}\\
        } \Else {
            $(p, i_{save}, d) = $ \textsf{Search}$(p, i, d, up, step_i)$ \textcolor{purple}{\Comment{Search dimension $i$ upwards}}\\
        }
        
        $BL.$\textsf{Insert}$(p, i_{save}, d)$ \textcolor{purple}{\Comment{Save boundary line}}\\
        $i = i_{save}$
    }
}
\caption{Boundary Identification Algorithm}
\label{algo:boundary_draw}
\end{algorithm}
\noindent \textbf{Coordinate Search.}
We describe the algorithm using coordinate search to find the boundary on a given two-dimensional plane $(k_i, k_d)$. In real-world UAVs, the boundary shape varies due to unpredictable external noises from the physical world and flight mode influence. For example, \Cref{fig:ArduPilot_boundary_identification_illustration} shows an approximately linear boundary as defined in \Cref{eq:linear_cond}. But the boundary in \Cref{fig:Noise_Illustration} turns into a drastically nonlinear sawtooth wave. Unlike conventional coordinate search aiming to find a particular data point \cite{wright2015coordinate}, our approach identifies the entire boundary line composed of a sequence of continuous data points. 

We start from an edge point (e.g., the upper-left corner) of the two-dimensional plane $(k_i, k_d)$. Then, we fix one dimension and search for another by a small step size to generate a new $(k_i, k_d)$ pair. We call the function \texttt{Search()} to perform this coordinate search. Note that the unknown gap can be extremely large, such that it distorts the slope direction of the theoretical boundary. We, therefore, adaptively search downwards and upwards in function \texttt{Search()} to ensure the discovery of the real boundary. After generating a candidate of the PID parameter configuration, we send it to the \textcolor{revisioncolor}{misbehavior validation} module to check if it is valid. The feedback information from the oracle is used to guide the future search direction. Once we find the first valid PID control parameter nearest the partition line, we save it to the result list. Different from conventional coordinate search, our algorithm continues the search after finding the first partition point. In the end, our algorithm terminates when it reaches the boundary of the two-dimensional plane $(k_i, k_d)$. Since it's an efficient coordinate search, the runtime overhead is $O(n)$, where $n$ denotes the number of steps along each dimension. 

Next, we explain the implementation details of our approach as shown in \Cref{algo:boundary_draw}. Lines 14-24 show the procedure of coordinate search. We iteratively search dimension $k_d$ with step size $step_d$. We then search dimension $k_i$ with the function \texttt{Search()} to find the valid PID control parameter that is nearest to the partition line and save it into the boundary line. 

\subsection{\textcolor{revisioncolor}{Misbehavior Validation}}
Our \textcolor{revisioncolor}{misbehavior validation} module classifies a PID parameter configuration as invalid if it leads to abnormal UAV flight behavior.
Accurate \textcolor{revisioncolor}{validation} of such misbehavior is crucial for identifying the valid range of PID parameters. Inspired by PGFuzz \cite{kim2021pgfuzz}, we used Metric Temporal Logic (MTL) 
\cite{koymans1990specifying} formula $\phi$ to precisely describe the specified behavior of UAVs from the document. 
If a UAV's flight behavior contradicts the MTL specification, the \textcolor{revisioncolor}{misbehavior validation} module will classify the configuration as invalid.
As \Cref{Verif_function} shows, by evaluating the flight trajectory $\omega$ against the MTL formula $\phi$,
We can verify whether the UAV's behavior conforms to the expected norms. Symbol $ \models $ indicates that trajectory $\omega$ conforms to MTL-formula $\phi$.
{\small
\begin{equation}
Verif(\omega, \phi) = 
\begin{cases} 
true & \text{if } \omega \models \phi \\
false & \text{otherwise}
\end{cases}
\label{Verif_function}
\end{equation}
}

\textcolor{revisioncolor}{For example, we use MTL formula $\square\{ (ALT_t<RTL\_ALT)\land(Mode_t=RTL)\rightarrow(ALT_{t-1}<ALT_t)\}$ to formally describe the expected UAV behavior during the ascending stage under ArduPilot \texttt{RTL} mode. $ALT_t$ and $ALT_{t-1}$ denote the altitude of the UAV at time steps $t$ and $t-1$, respectively. $RTL\_ALT$ denotes a threshold defined in the \texttt{RTL} mode. This MTL formula states that when the UAV operates in \texttt{RTL} mode, and its altitude is below $RTL\_ALT$, it should strictly increase its altitude. Any flight trajectory under ArduPilot \texttt{RTL} mode violating this MTL formula would be considered as UAV misbehaviors.}

The \textcolor{revisioncolor}{misbehavior validation} module comprises two main components: UAV simulation and \textcolor{revisioncolor}{offline oracle}. The UAV simulation operates with a set of PID control parameters and predefined flight commands. During the simulation, we record the UAV's flight log. Next, we \textcolor{revisioncolor}{leverage an offline oracle} to examine this log and check whether any misbehavior is observed.

\textcolor{revisioncolor}{We use the offline oracle because it can validate the entire flight trajectory and offer better validation accuracy than the online oracle. For flight modes that exhibit a long trajectory, the online oracle cannot accurately validate them. The reason is that the online oracle relies on a fixed-size sliding window, which cannot completely capture the UAV's flight trajectory. For example, ArduPilot's \texttt{Circle} mode requires the oracle to verify whether the flight trajectory exhibits a repeating circle. The trajectory is too long to fit into a fixed-size sliding window. Consequently, the online oracle often yields wrong results when given a partial UAV's flight trajectory. In contrast, the offline oracle can verify the MTL against the entire trajectory, which results in better accuracy. }

\section{Evaluation}
\label{sec:evaluation}
In this section, we evaluate the effectiveness of \prjName\ by answering the following research questions:

\noindent \textbf{RQ1: Can \prjName\ identify the accurate range of valid PID parameter configurations given a flight mode?}

RQ1 aims to evaluate the accuracy of the identified parameter range. 
In this context, we seek to determine whether there are any misconfigurations that our method fails to identify, as well as to verify if the PID configurations within the parameter region we've designated as invalid are indeed invalid.

\noindent \textbf{RQ2: How does \prjName\ perform in spotting individual invalid PID parameter configurations?}

RQ2 focuses on evaluating the efficacy of our method in discovering PID misconfigurations.
As misconfigurations provide developers with valuable information about flight control program vulnerabilities,
we want to evaluate our method's ability to detect misconfigurations.

\noindent \textbf{RQ3: Has our strategy for mitigating the influence of noise improved the accuracy of identified boundaries?}

RQ3 evaluates the effectiveness of the downward search strategy implemented within our coordinate search methodology. 
To mitigate the influence of noise, we incorporated this strategy into our search algorithm.
Our objective is to quantify the extent to which this search strategy enhances the overall performance of our approach.

\noindent \textbf{RQ4: How does the step size of the \prjName\ search algorithm affect the accuracy of boundary identification?}

The purpose of RQ4 is to examine the relationship between step size and boundary identification accuracy in the \prjName\ algorithm. The step size parameter is hypothesized to have a dual effect: smaller values may enhance accuracy by allowing for finer-grained boundary delineation, while potentially increasing computational time and reducing efficiency. This study systematically examines how accurately boundaries are identified with different step sizes to find the best balance between precision and computational efficiency in the \prjName\ algorithm.

{\color{revisioncolor} \noindent \textbf{RQ5: Does the sampling-based evaluation influence the \prjName's performance?}

RQ5 aims to assess the impact of sampling-based evaluation on the performance of \prjName. In our experiment, we sample the parameter space with a step size to evaluate the performance of \prjName\ due to the limited computational resources. This study investigates whether this sampling-based evaluation may affect the evaluation by comparing it against the exhaustive evaluation setting.}

{\color{revisioncolor} \noindent \textbf{RQ6: Is offline oracle more accurate than online oracle in UAV misbehavior validation?}

RQ6 evaluates the accuracy of an offline oracle over an online oracle in the validation of UAV misbehavior. The motivation for the offline oracle is that it improves the accuracy of validation by recording the flight states of all time steps, particularly in some flight modes (such as \texttt{AP:Circle}) where an accurate validation requires a long flight history. However, an online oracle with a sliding window cannot capture the entire flight trajectory and lead to an accurate validation. We compare the accuracy of the online and offline oracle. }

{\color{revisioncolor} \noindent \textbf{RQ7: Can other search algorithms (e.g., hill climbing and genetic algorithm) achieve comparable performance to coordinate search for boundary identification and misconfiguration detection?}

RQ7 investigates the application of other search algorithms (e.g., hill climbing and genetic algorithms) in \prjName. In our evaluation, we found coordinate search performed empirically the best among all three search algorithms in \prjName.  }

\subsection{Experimental Settings}
We conducted four studies to address our research questions (RQs). Each study is designed to provide insights into specific aspects of our method's performance and effectiveness. We conducted an
experiment on an openEuler release 20.03 LTS machine with 80 CPU cores (16 processors), \SI{251}{GB} RAM, and \SI{870}{GB} disk space.

\noindent \textbf{Subjects.} We studied two
popular flight control programs: ArduPilot (ArduCopter V4.4.1)\cite{ArduPilot-website} and PX4 (V1.15.0 a1cce7e) \cite{px4-website}. We evaluated our method in the SITL Simulator \cite{ArduPilot-sitl-website} of ArduPilot and the jMAVSim \cite{px4-jmavsim-website} of PX4 with default settings.
Our evaluation evaluated eight typical flight modes, encompassing both simple and complex flight tasks. From ArduPilot, we selected \texttt{RTL} \cite{ArduPilot_rtl_mode}, \texttt{Zigzag} \cite{zigzag-mode}, \texttt{Circle} \cite{circle-mode}, and \texttt{Brake} \cite{brake-mode}. From PX4, we chose \texttt{Orbit} \cite{px4_orbit_mode}, \texttt{Return} \cite{px4_retrun_mode}, \texttt{Land} \cite{px4_land_mode}, and \texttt{Hold} \cite{px4_hold_mode}.

\noindent \textbf{Experiment 0: Obtain Ground Truth.}
To evaluate the accuracy of these identified boundaries,
we obtain the \textcolor{revisioncolor}{ground truth benchmark} of the invalid PID parameter configurations.
The ground truth was obtained through a systematic exploration of the PID parameter space available in the document for each flight mode. Each PID parameter configuration was tested within its corresponding flight mode. Our \textcolor{revisioncolor}{misbehavior validation} module then classified each configuration as valid or invalid. This process, requiring approximately 7000 CPU hours, yielded \textcolor{revisioncolor}{a labeled benchmark} of PID parameter configurations for each flight mode.

\noindent \textbf{Experiment 1: Boundary Identification.} 
This experiment addresses RQ1. We utilize \prjName\ to identify the boundaries of invalid PID parameter regions. To evaluate the accuracy of these identified boundaries, we compare the invalid PID parameter configurations within these regions against the ground truth benchmark established in experiment 0. We choose this approach over comparing our method with other leading methods such as RVFuzzer and LGDFuzzer. \textcolor{revisioncolor} {We do not select RVFuzzer as a baseline because it is not open-sourced. Meanwhile, RVFuzzer's problem setting is not compatible with ours as it only outputs the value range for a single PID parameter, while our problem requires identifying a multi-dimensional PID parameter range. } \textcolor{revisioncolor}{
We do not select LGDFuzzer as our baseline, because in the context of a multi-dimensional parameter space, LGDFuzzer outputs the Cartesian product of these single-dimensional ranges. This results in a cuboid-like invalid parameter region, which does not match our polyhedron-like invalid parameter region in our problem definition. As a result, they are not suitable as our baselines.}

\noindent \textbf{Experiment 2: Misconfiguration Discovery.} This experiment addresses RQ2 by evaluating the effectiveness of our method in discovering PID parameter misconfigurations. We compare our method with PGFuzz, a state-of-the-art misconfiguration detection method. Both methods were run for 48 hours, and their performance was assessed based on the number of misconfigurations discovered. Our method utilizes identified boundaries of invalid PID parameter regions. Each parameter within these regions is tested under eight flight modes, and our \textcolor{revisioncolor}{misbehavior validation} module classifies it as either a misconfiguration or a valid configuration. PGFuzz uses a random search with simple guidance to find misconfigurations. For a fair comparison, we align PGFuzz's search space with ours, excluding parameters irrelevant to PID control modules. We also replace PGFuzz's online oracle with our offline \textcolor{revisioncolor}{misbehavior validator}. We ran the PGFuzz experiment three times to ensure robust results, given the variability of the random method.

\noindent \textbf{Experiment 3: Design Choice.}
This experiment addresses RQ3. We evaluate the impact of the downward search strategy on identifying valid PID parameter regions. We developed a variant method that omits the downward search during coordinate search and applied both the original and variant methods to eight flight modes. We then compared the boundaries identified by each method. This comparison assesses the downward search strategy's contribution to boundary accuracy.

\noindent \textbf{Experiment 4: Step Size Analysis.}
This experiment addresses RQ4, focusing on step size impact. We modified the original incremental step size by factors of 10x and 100x and then applied our search algorithm with these modified step sizes. We compare the resulting Miss Rate and Hit Rate to those obtained using the original step size. The comparison helps quantify the step size's influence on search accuracy and efficiency.

{\color{revisioncolor}
\noindent \textbf{Experiment 5: Sample-based Evaluation Influence.}
This experiment addresses RQ5. We exhaustively evaluated \prjName on all possible configurations of ArduPilot \texttt{RTL} mode. We then compared the miss rate and hit rate obtained from that exhaustive experiment setting with the rates from the sampling-based experiment setting.

\noindent\textbf{Experiment 6: Justification for Offline Oracle Design.}
This experiment addresses RQ6. We compared the accuracy of the offline oracle and online oracle to justify our selection of the offline oracle. We execute flight tasks across 8 flight modes under various PID configurations, and use online and offline oracles, respectively, to validate their behaviors. We calculated the accuracy of the different oracles by comparing the results of different oracle settings with human-verified results. 

\noindent\textbf{Experiment 7: Investigating Other Search Algorithms.}
This experiment addresses RQ7. We replaced the Coordinate Search component of \prjName\ with Hill Climbing (HC) and Genetic Algorithm (GA), creating RouthSearch-HC and RouthSearch-GA variants. These variants were applied to identify boundaries and detect misconfigurations with the same experiment settings of the original \prjName. We compare the miss rate, the hit rate, and the number of detected misconfigurations of two variants and the original \prjName.
}

\subsection{Evaluation Metrics} 
We use Miss Rate and Hit Rate to evaluate our method.
Let GT represent the set of all misconfigurations in the ground truth, and 
RS denotes the set of all misconfigurations generated by the \prjName\ method proposed in this study. The metrics employed in this paper are as follows:

\begin{description}
\item[MR.]
The Miss Rate (MR) quantifies the extent to which the boundaries identified by the proposed method fail to capture misconfigurations. A lower miss rate indicates a higher accuracy in identifying boundaries. It is defined as:
{\small
\[
MR = \frac{|GT - RS|}{|GT|}
\]
}

\item[HR.]
The Hit Rate (HR) measures the accuracy of the misconfigurations generated according to the boundaries identified by the proposed method. A higher hit rate signifies a more efficient generation of misconfigurations during the invalid configuration generation phase. It is defined as:
{\small
\[
HR = \frac{|RS \cap GT|}{|RS|}
\]
}

\end{description}

\subsection{Results: Experiment 1}
\label{subsec:rq1}

\begin{table}[!htbp]
    \footnotesize
  \centering
  \caption{Boundary identification accuracy of \prjName. This table shows the accuracy of the misconfigurations identified by the boundaries for different flight modes in ArduPilot (AP) and PX4 systems, measured in terms of total misconfigurations, identified misconfigurations, accurate detections, miss rate (MR), and hit rate (HR). Higher HR and lower MR indicate better detection capability.}
  \label{tab:boundary_identification_result_vs_groundtruth}
  \begin{tabular}{crrrrrr}
    \toprule
    \textbf{Mode} & \textbf{Total} & \textbf{Identified} & \textbf{Accurate} & \textbf{MR} & \textbf{HR} \\
    \midrule
        AP:Zigzag & 28,162 & 21,107 & 19,367 & 31.2\%  & 91.8\%\\
        AP:Brake & 36,584 & 33,357 & 32,902  & 10.1\% & 98.6\%\\      
        AP:RTL &  27,267 & 26,624 & 26,545 & 2.6\% & 99.7\% \\
        AP:Circle &  20,477 & 23,016 & 17,022 & 16.9\% & 74.0\% \\
        PX4:Orbit &  8,979 & 8,103 & 6,630 & 26.2\% & 81.8\% \\
        PX4:Return &  9,592 & 8,604 & 8,429 & 12.1\% & 98.0\% \\
        PX4:Land &  9,610 & 9,123 & 8,885 & 7.5\% & 97.4\%  \\
        PX4:Hold &  4,930 & 4,394 & 4,165 & 15.5\% & 94.8\% \\
        \hline
        Average & 18,200 & 16,791 & 15,493 & 15.3\% & 92.0\% \\
    \bottomrule
  \end{tabular}
\end{table}

\Cref{tab:boundary_identification_result_vs_groundtruth} shows the boundary identification performance of \prjName.  Column ``Total" shows the total ground truth misconfigurations. ``Identified" is the count of misconfigurations identified by \prjName. ``Accurate" is the count of correctly identified misconfigurations. ``MR" and ``HR" are the miss rate and hit rate, respectively.

The result shown in \Cref{tab:boundary_identification_result_vs_groundtruth} demonstrates our method's high effectiveness in boundary identification. The average hit rate (HR) of misconfigurations generated by our method is 92.0\%. This means that, on average, most of the PID parameter misconfigurations identified by \prjName\ are correct. The miss rate (MR), representing misconfigurations not detected by our method, averages 15.3\%. In other words, only a few PID parameter misconfigurations are not detected by our method. The high average Hit rate and the low average Miss Rate demonstrate the high accuracy of our boundary identification method.

Notably, in ArduPilot's \texttt{RTL} mode, RouthSearch achieves exceptional performance with a 99.7\% hit rate and only a 2.6\% miss rate, significantly surpassing the overall average performance of 92.0\% hit rate and 15.3\% miss rate across all flight modes. This indicates that our method is particularly effective in identifying PID parameter misconfigurations in this specific flight mode, with \texttt{RTL} mode showing the most precise detection capabilities among all tested scenarios.

The number of PID parameter misconfigurations varies across flight modes, suggesting the boundary shift of invalid PID parameter regions under different flight modes. This variation highlights the importance of considering flight modes when identifying PID parameter misconfigurations. A significant boundary shift is observed between different flight modes within ArduPilot: ArduPilot's \texttt{Brake} mode exhibits the highest number of invalid PID parameter configurations (36,584), while the \texttt{Circle} mode shows the lowest (20,477). This difference suggests that the \texttt{Brake} mode may be more sensitive to PID parameter misconfigurations compared to the \texttt{Circle} mode.

\begin{longfbox}
\textbf{Result 1:}
\prjName\ achieves an average 92.0\% hit rate and 15.3\% miss rate across various flight modes. This result indicates the effectiveness of our method in accurately identifying PID parameter misconfigurations.
\end{longfbox}
\subsubsection{Root Cause Analysis of Worst Case}

The performance of our method in the worst-case scenario is primarily influenced by two factors: the settling time of the PID control system \cite{ogata2020modern} and the \textcolor{revisioncolor}{misbehavior validator's} sensitivity.

\noindent \textbf{Long Settling Time.} ArduPilot's \texttt{Zigzag} mode exhibits an elevated MR (31.2\%), attributed to long settling times. The Routh-Hurwitz criteria guarantee stability as time approaches infinity but do not account for settling time, defined as the time for a system to stabilize \cite{ogata2020modern}. Small P values can lead to prolonged settling times, causing the system to slowly reach or even fail to reach the target \cite{meshram2012tuning,yeroglu2009new}. 
Consequently, our proposed method may overlook misconfigurations caused by these long settling times.

\begin{figure}[!ht]
    \centering
    \includegraphics[width=0.25\linewidth]{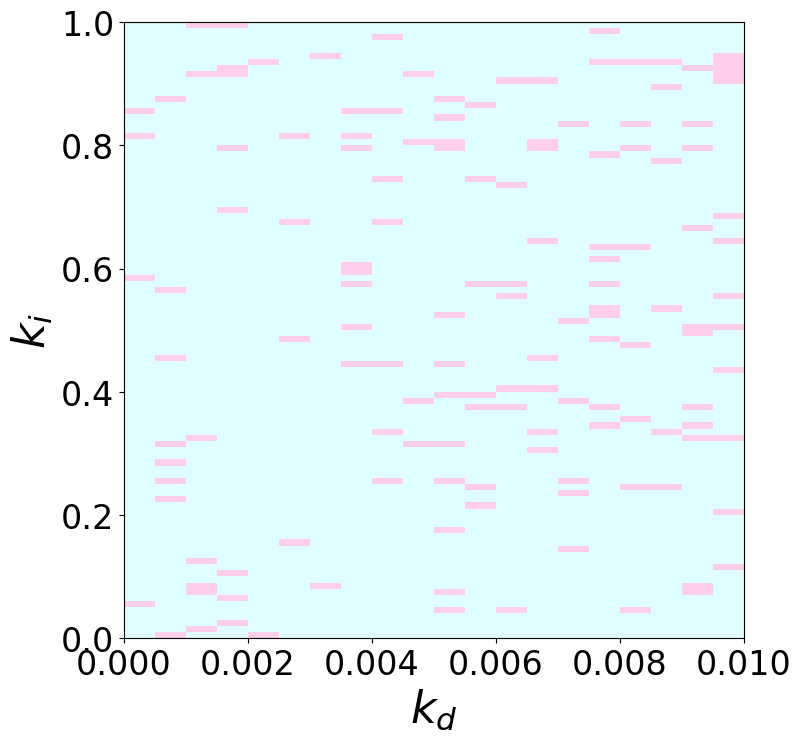}
    \Description{False negative reported by the \textcolor{revisioncolor}{misbehavior validator} under PX4 \texttt{Orbit} mode. 
    The \colorbox{ap-gt-pink}{pink} marks false positive misconfigurations caused by the detector's sensitivity to deviation.}
    \caption{False negative reported by the \textcolor{revisioncolor}{misbehavior validator} under PX4 \texttt{Orbit} mode. 
    The \colorbox{ap-gt-pink}{pink} marks false positive misconfigurations caused by the detector's sensitivity to deviation.}
    \label{fig:orbit_0.15_groundtruth}
\end{figure}

\noindent \textbf{Over-Sensitive \textcolor{revisioncolor}{Misbehavior Validator}.} 
Our method, when applied to PX4's \texttt{Orbit} mode and ArduPilot's \texttt{Circle} mode, exhibits a low hit rate (HR) of 81.8\% and 74.0\%. This discrepancy arises from the \textcolor{revisioncolor}{misbehavior validator's} sensitivity to the inherent deviations present in \texttt{Orbit} mode trajectories. While \texttt{Orbit} mode ideally maintains a circular flight path, factors like sensor inaccuracies introduce minor deviations that, although practically insignificant, impact the detector's ability to differentiate between valid and invalid trajectories. Consequently, valid PID parameter configurations are misclassified as invalid, as illustrated in \Cref{fig:orbit_0.15_groundtruth}, leading to an inaccurate boundary identified by our algorithm.

\subsection{Results: Experiment 2}
\Cref{tab:bugs-found-list} compares misconfiguration detection by \prjName\ and PGFuzz. Column ``\prjName'' is the number of \prjName's detections, and ``$PGFuzz_{\{1,2,3\}}$'' are PGFuzz detections over three runs.

\begin{table}[!htbp]
    \footnotesize
    \centering
    \caption{Comparison of number of misconfigurations discovered by \prjName and PGFuzz under three runs ($PGFuzz_{\{1,2,3\}}$) across flight modes in ArduPilot (AP) and PX4. Higher numbers indicate better performance.}
    \label{tab:bugs-found-list}
    \begin{tabular}{crrrr}
        \toprule
        \textbf{Mode} & \textbf{\prjName} & $\mathbf{PGFuzz_1}$ & $\mathbf{PGFuzz_2}$ & $\mathbf{PGFuzz_3}$ \\
        \midrule
        AP:Zigzag & 3,390 & 557 & 617 & 518\\

        AP:Brake &  11,820& 
        406 & 289 & 262 \\

        AP:RTL &  4,969 & 1,319  & 1,325  & 1,286 \\

        AP:Circle &  3,533  & 93 & 78 & 76 \\

        PX4:Orbit & 1,909 & 290 & 236 & 207\\

        PX4:Return & 1,092  & 299 & 278 & 256 \\

        PX4:Land & 2,037 &635 & 495 & 576 \\

        PX4:Hold & 2,070  & 305  & 193 & 171 \\

        \midrule
        Average & 3,853 & 488 & 439 & 419 \\
        \bottomrule
    \end{tabular}
\end{table}

As shown in \Cref{tab:bugs-found-list}, regarding misconfiguration generation, \prjName\ demonstrates superior effectiveness compared to the PGFuzz method. Our method discovered an average of 3,853 PID parameter misconfigurations, whereas PGFuzz detected a modest 449 misconfigurations (averaged across three experimental iterations). In the most successful instance, our method discovered 11,820 PID parameter misconfigurations in ArduPilot's \texttt{Brake} mode. In contrast, PGFuzz discovered an average of only 319 PID parameter misconfigurations, based on results from three iterations (406, 289, and 262 misconfigurations, respectively).

\begin{longfbox}
\textbf{Result 2:}
\prjName\ outperforms PGFuzz in PID parameter misconfiguration discovery, discovering 8.58 times more misconfigurations (3,853 vs 449 on average).
\end{longfbox}

\subsection{Results: Experiment 3}

\Cref{tab:ablation_test} shows the misconfiguration detection comparison between \prjName\ and \prjName-DSOff (Downward Search Off). Column ``Total'' represents the number of misconfigurations in ground truth, Column ``\prjName-DSOff'' and Column ``\prjName'' show the number of detected misconfigurations, Miss Rate, and Hit Rate for each.

\begin{table}[!htbp]
    \footnotesize
  \centering
  \captionsetup{labelfont={color=revisioncolor, small, sf}, textfont={color=revisioncolor, small, sf}}
  \caption{The total number of misconfigurations and accuracy discovered by \prjName\ and variant \prjName-DSOff  (Downward Search Off) for eight flight modes. Higher numbers, higher HR, and lower MR indicate better detection capability.}
  \begin{tabular}{c|r|rrr|rrr}
    \toprule
    
    \multirow{2}{*}{\textbf{Mode}} & \multirow{2}{*}{\textbf{Total}} & \multicolumn{3}{c|}{\textbf{\prjName-DSOff}} & \multicolumn{3}{c}{\textbf{\prjName}} \\
    & & \textbf{Number} & \textbf{MR} & {\color{revisioncolor} \textbf{HR}} & \textbf{Number} & \textbf{MR} & {\color{revisioncolor} \textbf{HR}} \\
    
    \midrule
        AP:Zigzag & 28,162 & 531 & 98.1\% & {\color{revisioncolor} 97.3\%} & 19,367  & 31.2\% & {\color{revisioncolor} 91.8\%}\\
        AP:Brake & 36,584 & 32,515 & 11.1\% & {\color{revisioncolor} 98.2\%} & 32,902 & 10.1\% & {\color{revisioncolor} 98.6\%}\\
        AP:RTL & 27,267 & 26,531 & 2.7\% & {\color{revisioncolor} 99.7\%} & 26,545 & 2.6\% & {\color{revisioncolor} 99.7\%}\\
        AP:Circle & 20,477 & 15,678 & 23.4\% & {\color{revisioncolor} 93.1\%} & 17,022 & 16.9\% & {\color{revisioncolor} 74.0\%}\\
        PX4:Orbit & 8,979 & 3,779 & 57.9\% & {\color{revisioncolor} 99.9\%} & 6,630 & 26.2\% & {\color{revisioncolor} 81.8\%}\\
        PX4:Return & 9,592 & 5,340 & 44.3\% & {\color{revisioncolor} 99.7\%} & 8,429 & 12.1\% & {\color{revisioncolor} 98.0\%}\\
        PX4:Land & 9,610 & 6,737 & 29.9\% & {\color{revisioncolor} 98.7\%} & 8,885 & 7.5\% & {\color{revisioncolor} 97.4\%}\\
        PX4:Hold & 4,930 & 3,175 & 35.6\% & {\color{revisioncolor} 99.5\%} & 4,165 & 15.5\% & {\color{revisioncolor} 94.8\%} \\
        \hline
        Average & 18,200 & 11,786 &  37.9\% & {\color{revisioncolor} 98.3\%} & 15,493 & 15.3\% & {\color{revisioncolor} 92.0\%}\\
    \bottomrule
  \end{tabular}
  \label{tab:ablation_test}
\end{table}

As shown in \Cref{tab:ablation_test}, \prjName\ achieves significantly better results than \prjName-DSOff in terms of the number of identified misconfigurations and the miss rate. On average, \prjName\ identifies 15,493 misconfigurations and a 15.3\% miss rate. \prjName-DSOff identifies 11,786 misconfigurations and 37.9\% miss rate. That means the downward search strategy can significantly decrease the miss rate (from 37.9\% to 15.3\%). 
{\color{revisioncolor} RouthSearch-DSOff's hit rate slightly improved because the RouthSearch-DSOff doesn't explore the noise area, where invalid and valid configurations are mixed.}
In other words, the downward search strategy is effective in identifying configuration boundaries. 

\begin{longfbox}
\textbf{Result 3:} 
The downward search strategy is effective in identifying configuration boundaries, decreasing the miss rate from 37.9\% to 15.3\%.
\end{longfbox}

\subsection{Results: Experiment 4}
% add runtime information
\begin{table}[htbp]
    \footnotesize
  \centering
  \caption{The influence of step size in algorithm 1 on the misconfiguration detection and boundary identification accuracy of \prjName. The step size ranges from 1x (finest) to 100x (coarsest). Results are measured in terms of total misconfigurations, identified misconfigurations, accurate detections, miss rate (MR), and hit rate (HR). Higher HR and lower MR indicate better performance.}
  \begin{tabular}{crrrrr}
    \toprule
    \textbf{Step Size} & \textbf{Total} & \textbf{Identified} & \textbf{Accurate} & \textbf{MR} & \textbf{HR} \\
    \midrule
    1x & 374,578 & 330,727 & 325,300 & 13.2\% & 98.4\%\\
    10x & 36,584 & 33,357 & 32,902  & 10.1\% & 98.6\%\\
    100x & 4,012 & 3,812 & 3,673 & 8.4\% & 96.4\%\\
    \bottomrule
  \end{tabular}
  \label{tab:step_size_influence}
\end{table}

\Cref{tab:step_size_influence} examines how step size affects detection performance.  The ``Total'' column represents the overall misconfiguration ground truth.  ``Identified'' indicates the count of misconfigurations \prjName\ invalidates.  ``Accurate'' reflects the number of correctly detected invalid misconfigurations.  ``MR'' and ``HR'' present the Miss Rate and Hit Rate evaluation metrics.

The results reveal that the 10x scale step size achieves the highest hit rate (98.6\%), while the 100x scale step size yields the lowest (96.4\%). However, the difference between these extremes is merely 2.2\%. Regarding miss rate, the 100x scale step size performs best (8.4\%), with the 1x scale step size showing the poorest performance (13.2\%). The difference here is 4.8\%. These relatively small variations indicate that step size has a limited impact on the method's overall performance.

Considering both accuracy and computational efficiency, we empirically determined that a 10x scale step size provided the optimal trade-off for our experiments. This choice balances the high hit rate (98.6\%) with a reasonably low miss rate (10.1\%), while significantly reducing the configurations to be tested compared to the 1x scale.

\begin{longfbox}
\textbf{Result 4:} 
The maximum differences in the MR and HR across various step sizes were 4.8\% and 2.2\% respectively. These small variations indicate that our method's performance is relatively insensitive to step size, with the 10x scale step size offering the most balanced selection.
\end{longfbox}

{\color{revisioncolor} \subsection{Results: Experiment 5}
\Cref{tab:ground_truth_comparison} presents a comparison between computing HR and MR using exhaustive and sampling evaluation.  ``Exhaustive'' (evaluation over all possible configurations) and ``Sampling-based'' (evaluation over sampled configurations). Column ``CPU hours'' means total runtime.}

\begin{table}[htbp]
    \footnotesize
  \centering
  {\color{revisioncolor}
  \captionsetup{labelfont={color=revisioncolor, sf, small}, textfont={color=revisioncolor, sf, small}}
  \caption{Exhaustive evaluation vs Sampling-based evaluation; Exhaustive means evaluation over all possible configurations; Sampling-based means evaluation over sampled configurations. CPU hours mean total runtime.}
  \label{tab:ground_truth_comparison}
  \begin{tabular}{c|c|r|r|r}
    \toprule
    \textbf{Mode} & \textbf{Experimental Setting} & \textbf{MR} & \textbf{HR} & \textbf{CPU hours} \\
    \midrule
    \multirow{2}{*}{AP: RTL} & Exhaustive & 6.7\% & 98.9\% & 9,067 \\
     & Sampling-based & 2.6\% & 99.7\% & 907 \\
    \bottomrule
  \end{tabular}}
\end{table}

{\color{revisioncolor} As \Cref{tab:ground_truth_comparison} shows, we found that the difference in MR and HR is marginal, which indicates that the sampling-based evaluation is reliable. The variance in MR (6.7\% vs 2.6\%) and HR (98.9\% vs 99.7\%) is minor. Meanwhile, our sampling-based evaluation requires only one-tenth total runtime compared with exhaustive evaluation. }

\begin{longfbox}
 {\color{revisioncolor}
\textbf{Result 5:} 
Our sampling-based evaluation is reliable with minimal influence on the MR and HR, while also requiring only one-tenth of the total runtime compared with exhaustive evaluation.}
\end{longfbox}

{\color{revisioncolor} \subsection{Results: Experiment 6}
\Cref{tab:oracle_performance_comparison} compares agreement rate with expert classification (\%) of ``Online" and ``Offline" Oracles. Higher percentages indicate better performance.}

\begin{table}[htbp]
    \footnotesize
  \centering
  {\color{revisioncolor}
  \captionsetup{labelfont={color=revisioncolor, small, sf}, textfont={color=revisioncolor, small, sf}}
  \caption{Online/Offline method Detection Accuracy Comparison (evaluated by agreement rate with human-verified results). }
  \label{tab:oracle_performance_comparison}
  \begin{tabular}{c|r|r|c|r|r}
    \toprule
    \textbf{Mode} & \textbf{Online} & \textbf{Offline} & \textbf{Mode} & \textbf{Online} & \textbf{Offline}\\
    \midrule
    AP:Brake & 96.5\% & 99.4\% & PX4:Hold & 100\% & 100\% \\
    AP:Circle & 72.7\% & 96.5\% & PX4:Land & 100\% & 100\% \\
    AP:RTL & 99.7\% & 99.7\% & PX4:Return & 100\% & 100\% \\
    AP:Zigzag & 97.1\% & 97.1\% &  PX4:Orbit & 28.5\% & 97.3\% \\
    \bottomrule
  \end{tabular}}
\end{table}

{\color{revisioncolor} \Cref{tab:oracle_performance_comparison} shows that the offline oracle generally outperforms the online oracle for UAV misbehavior validation, due to its use of the entire trajectory context. 
 For the long-term \texttt{PX4:Orbit} mode, the offline oracle is significantly better, reaching 97.3\% agreement compared to online's 28.5\% (68.8\% higher). This gap in \texttt{PX4:Orbit}, a long-duration mode, highlights the online oracle's limited window size, failing to capture the long-term context needed for complex misbehavior validation. The offline oracle overcomes this by using the entire trajectory. For short-term flight mode, the offline oracle's accuracy is also on par with the online oracles' accuracy. For instance, both online and offline oracles achieve high agreement (96.5\%-100\%) in modes like \texttt{AP:Brake} and \texttt{PX4:Hold/Land/Return}. In conclusion, \Cref{tab:oracle_performance_comparison} empirically justifies offline oracle for RQ6, as it better detects UAV misbehavior, especially those requiring the entire trajectory context other than limited slide windows. }

\begin{longfbox}
{\color{revisioncolor}
\textbf{Result 6:}
The offline oracle outperforms the online oracle for UAV misbehavior validation, especially for long-term patterns (\texttt{PX4:Orbit} 97.3\% vs 28.5\% agreement rate, 68.8\% higher). For short-term patterns, both methods show similar high performance (96.5–100\%).
}
\end{longfbox}

{\color{revisioncolor} \subsection{Results: Experiment 7}
\Cref{tab:routh_search_comparison} compares the performance of RouthSearch variants: Coordinate Search (CS), Hill Climbing (HC), and Genetic Algorithm (GA).  The columns show Miss Rate (MR), Hit Rate (HR), and the number of misconfigurations identified ('number') for each method.
}

\begin{table}[htbp]
  \centering
  \scriptsize
  {\color{revisioncolor}
  \captionsetup{labelfont={color=revisioncolor, small, sf}, textfont={color=revisioncolor, small, sf}}
  \caption{Comparison performance between RouthSearch-CS(Routh  Search with Coordinate Search), RouthSearch-HC(Routh Search with Hill Climbing) and RouthSearch-GA(Routh   Search with Genetic Algorithm). Higher numbers, higher HR, and lower MR indicate better detection capability.}
  \label{tab:routh_search_comparison}
  \setlength{\tabcolsep}{7pt}
  \renewcommand{\arraystretch}{1.2}
  \begin{tabular}{c|c|c|c|c|c|c|c|c|c}
    \toprule
    \multirow{2}{*}{\textbf{Mode}} & \multicolumn{3}{c|}{\textbf{RouthSearch-CS}} & \multicolumn{3}{c|}{\textbf{RouthSearch-HC}} & \multicolumn{3}{c}{\textbf{RouthSearch-GA}} \\
    & \textbf{MR} & \textbf{HR} & \textbf{number} & \textbf{MR} & \textbf{HR} & \textbf{number} & \textbf{MR} & \textbf{HR} & \textbf{number} \\
    \midrule
    AP:Zigzag & 31.2\% & 91.8\% & 3390 & 98.3\% & 25.1\% & 481 & 93.0\% & 20.2\% & 1928 \\
    AP:Brake & 10.1\% & 98.6\% & 11,820 & 93.5\% & 49.3\% & 2373 & 89.9\% & 35.5\% & 3688 \\
    AP:RTL & 2.6\% & 99.7\% & 4,969 & 95.0\% & 37.9\% & 1379 & 88.4\% & 30.3\% & 3204 \\
    AP:Circle & 16.9\% & 74.0\% & 3,533 & 95.4\% & 45.6\% & 934 & 90.2\% & 19.1\% & 1874 \\
    PX4:Orbit & 26.2\% & 81.8\% & 1,909 & 98.0\% & 22.8\% & 173 & 96.4\% & 42.2\% & 319 \\
    PX4:Return & 12.1\% & 98.0\% & 1,092 & 98.3\% & 37.6\% & 162 & 96.4\% & 40.3\% & 345 \\
    PX4:Land & 7.5\% & 97.4\% & 2,037 & 96.3\% & 40.8\% & 358 & 95.8\% & 39.3\% & 401 \\
    PX4:Hold & 15.5\% & 94.8\% & 2,070 & 96.8\% & 21.9\% & 158 & 94.5\% & 29.8\% & 270 \\
    \hline
    Average & 13.0\% & 92.0\% & 3853 & 96.5\% & 35.1\% & 752 & 93.1\% & 32.1\% & 1504 \\
    \bottomrule
  \end{tabular}}
\end{table}

{\color{revisioncolor}  \Cref{tab:routh_search_comparison} demonstrates that Coordinate Search(CS) is better than Hill Climbing (HC) and Genetic Algorithms (GA) in \prjName. CS only requires binary feedback to decide on whether to search upward or downward, which aligns perfectly with the binary feedback, achieving near-perfect precision (e.g., 2.6\% MR and 99.7\% HR in \texttt{AP:RTL}) and 2.5 times higher misconfiguration counts.  In contrast, HC exhibits excessively high miss rates (e.g., 98.3\% MR in \texttt{AP:Zigzag}) and low hit rates (e.g., 21.9\% HR in \texttt{PX4:Hold}). This limitation arises because hill climbing lacks quantitative feedback, degenerating into a local random search that produces narrow, inaccurate boundaries.
GA also exhibits high miss rates and low hit rates. Although GA is slightly better than HC (for example, 89. 9\% vs. 93. 5\% in AP: Brake) through a greater exploration, but remains trapped in undirected randomness due to missing fitness-guided refinement, resulting in persistently low HR (19.1–42.2\%) and suboptimal misconfiguration detection (max: 3,688 vs. 11,820 for \prjName-CS).}

\begin{longfbox}
{\color{revisioncolor}
\textbf{Result 7:}
Coordinate search significantly surpasses hill climbing and genetic algorithms, achieving superior HR and MR and detecting 2.5-5 times more misconfigurations on average. This result validates the selection of coordinate search due to its effective use of binary feedback.
 }
\end{longfbox}
\subsection{Case Study}
By utilizing our method in the ArduPilot and PX4 flight control programs,
We obtained many failed cases and spotted several bugs.
Here are some bug cases we have found.

\subsubsection{\textcolor{revisioncolor}{Bug 1}}
\begin{figure}[!ht]
  \centering
  \includegraphics[width=0.25\linewidth]{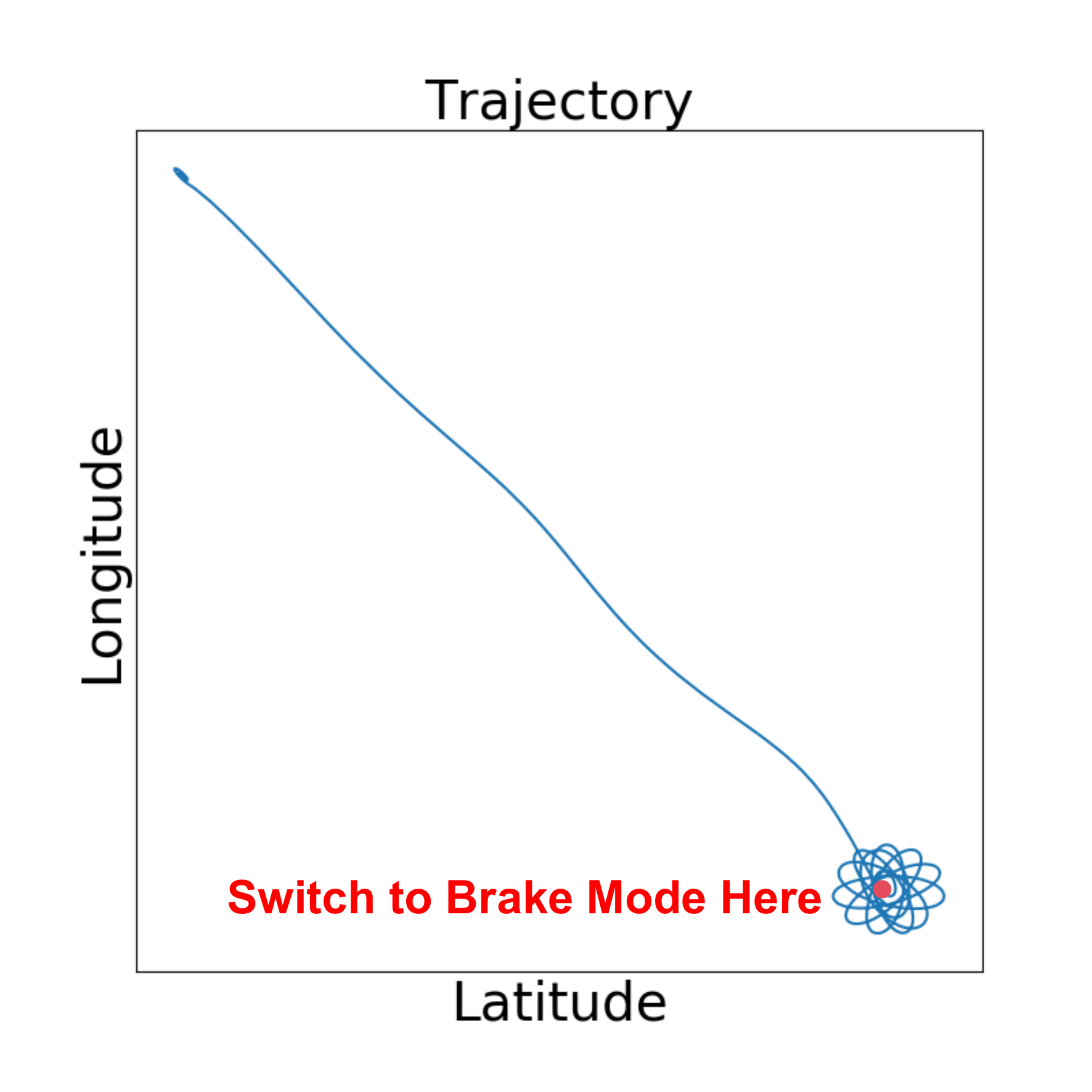}
  \Description{UAV under ArduPilot \texttt{Brake} mode fails to stop at the destination marked in \textbf{\textcolor{red}{red}} and gets stuck in circles endlessly due to a logic bug in the flight control program.}
  \caption{UAV under ArduPilot \texttt{Brake} mode fails to stop at the destination marked in \textbf{\textcolor{red}{red}} and gets stuck in circles endlessly due to a logic bug in the flight control program.}
  \label{fig:brake_mode_abnormal_case}
\end{figure}

As the ArduPilot document said, the \texttt{Brake} mode stops the vehicle as soon as possible. Once invoked, this mode does not accept any input from the pilot. 
We have found that PID misconfigurations could cause the \texttt{Brake} mode to fail to stop the vehicle 
and move around. After misconfiguring the PID parameters of the ArduPilot position control sub-module into $(0.1, 0.8, 0.8)$, we commanded the UAV to fly for a certain distance and then issued a command to enter \texttt{Brake} mode. The UAV attempted to stop quickly and regain its balance. However, due to the misconfiguration, the more the flight control program attempts to utilize the PID control module to quickly bring the UAV to a stop, the more the unstable PID control module fails to achieve the objective. This results in the UAV continuously rotating, as shown in \Cref{fig:brake_mode_abnormal_case}.

To make it even worse, due to the refusal of \texttt{Brake} mode to accept commands sent by the sticks of the transmitter.
Thus, the user can't stop the UAV manually while the UAV is in a perilous state.
As \Cref{code:try_to_stop} shows, ArduPilot is trying to stop the UAV by setting a zero velocity as the target to the PID control module. While the UAV is unstable because of the misconfiguration, the velocity is always greater than zero, which results in a rotating UAV that never stops. This issue represents a logical flaw in the system, where there is inadequate error handling for scenarios in which the speed fails to reach zero.

\begin{lstlisting}[float=h,style=cpp,xleftmargin=10pt,caption={The logical bug in \texttt{Brake} mode of ArduPilot. The ArduPilot tries to stop the UAV by setting a zero velocity as the target to the PID control module, but the unstable UAV fails to achieve that due to the misconfiguration. That results in the UAV continuously rotating, because the flight control program forgets to handle the error case when the function call \texttt{set\_target\_velocity(0)} fails. },label={code:try_to_stop}, basicstyle=\scriptsize\ttfamily]
void brake_mode() {
    while(true) {
        pos_pid_control.set_target_velocity(0);
        /*Logical Bug: Missed error handling.*/
        sleep(0.1);
    }
}
\end{lstlisting}
  
\subsubsection{\textcolor{revisioncolor}{Bug 2}}
\label{subsec:Motivated_Example}
\begin{figure}[!htbp]
    \centering
    \includegraphics[width=0.25\linewidth]{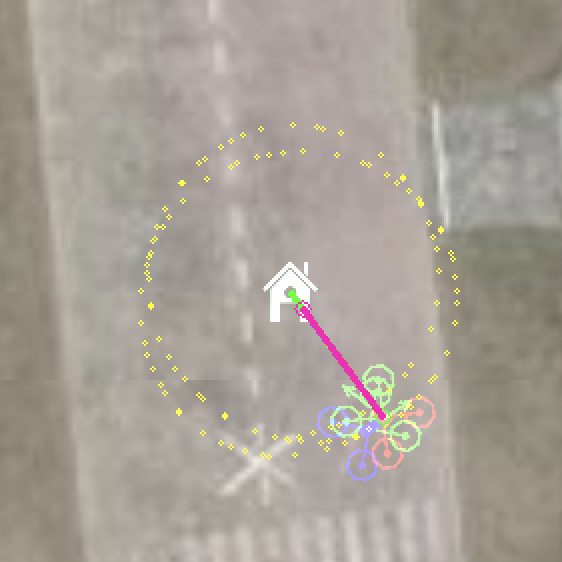}
    \Description{Flight trajectory (marked in \textbf{\textcolor{ap-traj-yellow}{yellow}}) of a misconfigured UAV (marked in \textbf{\textcolor{ap-traj-green}{green}}) under \texttt{RTL} mode. The UAV, stuck in a circle around the home location and failed to land, was caused by misconfigured PID parameters. }
    \caption{Flight trajectory (marked in \textbf{\textcolor{ap-traj-yellow}{yellow}}) of a misconfigured UAV (marked in \textbf{\textcolor{ap-traj-green}{green}}) under \texttt{RTL} mode. The UAV, stuck in a circle around the home location and failed to land, was caused by misconfigured PID parameters. }
    \label{fig:copter-ossilization}
\end{figure}

\begin{lstlisting}[float=h,style=cpp,xleftmargin=10pt,caption={Logical bug in ArduPilot \texttt{RTL} mode. ArduPilot checks the UAV's approach to the descent location by calculating the distance to it. Misconfigured PID parameters cause the UAV to circle continuously, preventing it from getting close enough to reduce the distance under threshold \texttt{radius}. In that case, \texttt{state\_complete} is always false. The flight control program forgets to handle the error case. Therefore, the landing action is not invoked, and the UAV fails to land at the home location.},label={code:switch_to_land}, basicstyle=\scriptsize\ttfamily]
// Check if the UAV reaches the destination
bool reached_dst() {
    dist = compute_dist(curr_loc, dst);
    if(dist < radius) 
        return true;
    return false;
}

// Attempt to land after the UAV reaches the destination
void switch_to_land() {
    state_complete = reached_dst();
    if(state_complete)
        perform_land();
    /* Logical Bug: Missed error handling. */
}
\end{lstlisting}

\texttt{RTL (Return-To-Launch)} is an ArduPilot flight mode that commands the UAV to fly towards its home location and land after hovering above for a while. 
A logic bug occurs in \texttt{RTL} mode when users set the PID parameters with a value that would cause the UAV to misbehave. Specifically, users set the PID parameters of the position control sub-module as $(0.1, 0.8, 0.4)$, all of which fall within the value range of the PID control parameter in the official document of ArduPilot. However, these PID parameters accidentally lead to an unstable control state of the UAV under \texttt{RTL} mode. As a result, the UAV kept circling around the home location without reaching a suitable descending spot. ~\Cref{fig:copter-ossilization} shows the flight trajectory (marked in \textcolor{ap-traj-yellow}{yellow}) of the UAV (\textcolor{ap-traj-green}{green}). Given that the goal of \texttt{RTL} mode is to perform a landing at the home location, the misconfigured UAV under \texttt{RTL} mode is trapped in a circle and fails to accomplish the landing task.  

\Cref{code:switch_to_land} shows the logical bug in ArduPilot. In Line 2, the function \texttt{reached\_dst()} checks if the UAV reaches a suitable descent location by measuring the distance between the UAV's current position and the descent location. The distance is shown in \textbf{\textcolor{ap-traj-purple}{purple}} in \Cref{fig:copter-ossilization}. Due to misconfigured PID control parameters, the UAV is stuck in a circle and fails to reduce the distance under the threshold \texttt{radius}. In Line 12, the \texttt{state\_complete} is always false. The landing action in Line 13 is, therefore, not invoked. Ultimately, the UAV hovers in a circle and fails to land at the home location. 
It's worth noting that we have already brought this issue to the attention of the developers, who have acknowledged the bug.

\subsubsection{\textcolor{revisioncolor}{Bug 3}}
\begin{figure}[!htbp]
  \centering
    \centering
    \includegraphics[width=0.25\linewidth]{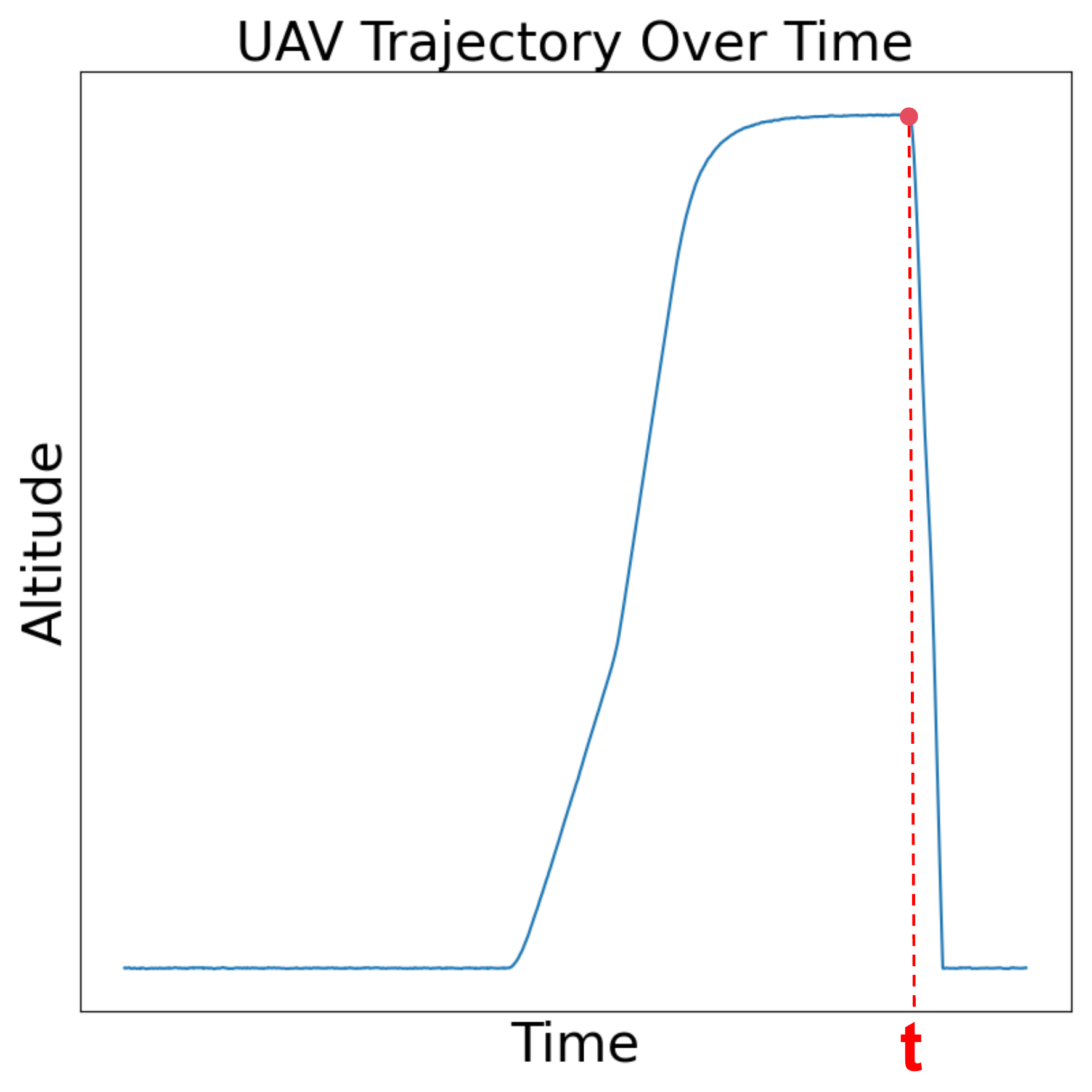} 
   \Description{UAV trajectory of PX4 under \texttt{Hold} mode. UAV crashes after \textbf{\textcolor{red}{time t}} when a user misconfigures a PID parameter.}
  \caption{UAV trajectory of PX4 under \texttt{Hold} mode. UAV crashes after \textbf{\textcolor{red}{time t}} when a user misconfigures a PID parameter.}
\label{fig:hold_mode_abnormal_case}
\end{figure}

As the PX4 document said, the \texttt{Hold} mode causes the vehicle to stop and maintain its current position and altitude \cite{px4_hold_mode}.
As \Cref{fig:hold_mode_abnormal_case} shows,
we have found that PID misconfigurations could cause the \texttt{Hold} mode to fail to hold the UAV's position and attitude
and lead the UAV to crash to the ground. After commanding the UAV to take off to a certain altitude, we misconfigured the PID parameters of the PX4 attitude control submodule into $(0.01, 0.1, 0.0005)$ and then issued a command to enter \texttt{Hold} mode.
As \Cref{fig:hold_mode_abnormal_case} shows,
once we have misconfigured PID parameters,
the UAV becomes unstable and quickly crashes to the ground due to the loss of balance.

{\color{revisioncolor}\subsubsection{Discussion}
In three 48-hour experiment runs,  \prjName\ exclusively discovered Bugs 1 and 2, which PGFuzz failed to discover. While PGFuzz occasionally found Bug 3 in only two of the three runs and required $8.86$ hours for detection in those instances, \prjName\ consistently identified the same bug in just 1.5 minutes. Furthermore, \prjName discovered $2,002$ PID misconfigurations, compared to the $7$ found by PGFuzz. The random and isolated parameter mutation of PGFuzz hindered the detection of multidimensional trigger conditions (required by Bugs 1 and 2). In contrast, \prjName's principled PID parameter analysis and theoretical guidance enabled systematic misconfiguration detection. Bug 3, requiring a single-dimensional trigger (a low value of $p$), can be occasionally found by PGFuzz, but \prjName\ can detect it more consistently and efficiently.}

\section{Limitations and Discussion}
\label{sec:limitation_and_discussion}

\begin{description}[leftmargin=0cm]
\item[Handling Misconfigurations Caused by Long Settling Time. ] 
The settling time problem occurs when both the $p$ value and $i$ value are simultaneously small, the PID control module fails to achieve the desired output, and causes the system under control to move too slowly to reach the target.
That results in PID misconfigurations that our method may overlook.
A candidate fix is to examine the configurations where both the $p$ value and $i$ value are simultaneously small in pre-processing.

\item[\textcolor{revisioncolor}{Misbehavior Validator's} Sensitivity.] 
The detector's over-sensitivity affects boundary identification accuracy. It sometimes misclassifies valid PID configurations as invalid. This can lead to inaccurate boundary identification in corner cases. A potential solution is to use multiple \textcolor{revisioncolor}{misbehavior validators} and perform majority voting. This approach could mitigate the impact of deviations on the detector's output.

\item[Trade-off between Accuracy and Efficiency.]  
A large search step size in boundary identification improves \prjName's efficiency but may reduce boundary accuracy. Conversely, a small step size increases accuracy at the expense of time.  Selecting an appropriate step size that balances the trade-off between these requirements should be considered in practice.
\end{description}

\section{Related Works}
\label{sec:related_works}

\noindent \textbf{PID Parameters Optimization.}
In the field of control theory, researchers have long focused on strategies for selecting optimal PID control parameters.
Classical approaches, including Ziegler-Nichols \cite{meshram2012tuning}, Åström-Hägglund \cite{yeroglu2009new}, and Cohen-Coon \cite{joseph2017cohen}, are widely employed to determine suitable PID control parameters within specified ranges for various control systems. Building upon these classical methods, more sophisticated techniques have emerged to address diverse control challenges. Computational intelligence approaches, such as genetic algorithms \citep{jayachitra2014genetic, ibrahim2019optimal} and particle swarm optimization \citep{herlambang2019particle, kashyap2021particle}, offer robust tuning strategies for complex systems where traditional methods may fall short. 
Data-driven tuning approaches \citep{jesawada2022model, sonmez2024reinforcement} generate tuning strategies through learning processes, combining system models and operational data to optimize control parameters.
Recent advancements \citep{ghamari2023lyapunov, javadian2023evolutionary, purandare2023self} focus on robust and adaptive tuning strategies to manage system uncertainties and time-varying dynamics, addressing limitations of static tuning methods.

Based on these theories, software developers have implemented numerous tools for fine-tuning PID control parameters across various applications, such as the \texttt{AutoTune mode} \cite{autotune-mode} in ArduPilot. Nevertheless, these tools may pose challenges for normal users lacking substantial expertise in tuning control parameters \cite{ardupilot2023payload}. 

\noindent \textbf{UAV Parameter Misconfiguration Detection.}
The misconfiguration of parameters in UAVs frequently results in abnormal flight behavior, jeopardizing flight safety. In recent years, ensuring the flight safety of unmanned aerial vehicles (UAVs) has become a focal point of considerable attention from researchers \citep{wang2021exploratory, di2023automated, park2023scvmon}. 

Fuzz testing has been adapted for UAVs to uncover bugs arising from misconfigured parameters. Unlike traditional software fuzzing that detects crashes or memory corruption, bugs in UAV systems typically manifest in distinct error types \citep{kim2022reverse, schiller2023drone}. Recent UAV fuzzing methods address this using policies based on physical behavior to guide testing \citep{kim2022robofuzz, choi2020cyber}. PGFuzz \cite{kim2021pgfuzz}, for example, implements temporal logic policies to guide the search for erroneous UAV parameters. These policy-driven fuzzing methods excel in uncovering bugs that conventional methods may overlook. Concurrently, learning-guided fuzzing techniques, used in testing Cyber-Physical Systems \citep{zhang2021figcps, chen2019learning}, are also used to identify UAV parameter bugs. LGDFuzzer \cite{han2022control}, for instance, employs a learned model that effectively guides the search.

\section{Conclusion}
\label{sec:conclusion}
This paper proposes \prjName, a novel method leveraging Routh-Hurwitz stability and coordinate search to accurately determine the valid 3D ranges for critical PID parameters in the UAV flight control program. \prjName\ identifies boundaries between valid and invalid PID parameter configurations, helping prevent users from misconfiguring PID parameters during flight. Evaluated on ArduPilot and PX4 across 8 flight modes, it achieved 92.0\% accuracy, detected ~3,850 misconfigurations per 48 hours (significantly outperforming prior work), and uncovered 3 bugs.

\section{Data Availability}
We provide our tool \prjName, an implementation of the boundary identification module and \textcolor{revisioncolor}{misbehavior validation} module in web at \href{https://github.com/SciC0d3m4xOfW/RouthSearch}{https://github.com/SciC0d3m4xOfW/RouthSearch}. 

\begin{acks}
This work was supported by the National Key R\&D Program of China under Grant \\ No. 2022YFB4501803. This work was also supported by the National Natural Science Foundation of China (Grant No. 62472100).
\end{acks}
\clearpage

\bibliographystyle{ACM-Reference-Format}
\bibliography{references.bib}
\end{document}